\documentstyle{mn}
\input{psfig.sty}
\newcommand{\ndisk}	{\mbox{\boldmath$n_{\rm disk}$}}

\newcommand{\Jdot}	{\dot{J}}
\newcommand{\be}   	{\begin{equation}}
\newcommand{\ee}   	{\end{equation}}
\newcommand{\bea}	{\begin{eqnarray}}
\newcommand{\eea}	{\end{eqnarray}}
\newcommand{\bfT}       {\mbox{\boldmath$T$}}
\newcommand{\bfOm}      {\mbox{\boldmath$\Omega$}}
\newcommand{\bfm}       {\mbox{\boldmath$m$}}
\newcommand{\bfTone}    {\mbox{\boldmath$T_{1}$}}
\newcommand{\bfTtwo}    {\mbox{\boldmath$T_{2}$}}
\newcommand{\bfJ}       {\mbox{\boldmath$J$}}

\newcommand{\rBtwelve}  {\left( \frac{10^{12} \rm \, G}{B} \right)}
\newcommand{\Bnine}     {\left( \frac{B}{10^{9} \rm \, G} \right)}

\newcommand{\foh}       {\left( \frac{f}{100 \rm \, Hz} \right)}
\newcommand{\rfoh}      {\left( \frac{100 \rm \, Hz}{f} \right)}
\newcommand{\fHz}       {\left( \frac{f}{1 \, \rm Hz} \right)}

\newcommand{\taue}      {\tau_{\rm e}}
\newcommand{\taua}      {\tau_{\rm a}}
\newcommand{\taup}      {\tau_{\rm pump}}
\newcommand{\taud}      {\tau_{\rm d}}
\newcommand{\tauehat}   {\hat{\tau}_{\rm e}}
\newcommand{\thetam}    {\theta_{\rm max}}
\newcommand{\Pfp}       {P_{\rm fp}}
\newcommand{\thns}      {\theta_{\rm ns}}
\newcommand{\thetad}    {\dot{\theta}}
\newcommand{\phid}      {\dot{\phi}}
\newcommand{\phidd}     {\ddot{\phi}}
\newcommand{\psid}      {\dot{\psi}}
\newcommand{\td}	{\dot{\theta}}
\newcommand{\DId}       {\Delta I_{\rm d}}
\newcommand{\DIOm}      {\Delta I_{\Omega}}
\newcommand{\ronekpc}   {\left( \frac{1 \rm \, kpc}{r} \right)}

\newcommand{\Io}        {I_{0}}
\newcommand{\roneyear}  {\left( \frac{1 \rm \, yr}{\taue} \right)}
\newcommand{\Medd}      {\left( \frac{\dot{M}}{\dot{M}_{\rm E}} \right)}
\newcommand{\st}        {\sin \theta}
\newcommand{\ct}        {\cos \theta}
\newcommand{\sst}       {\sin^{2} \theta}
\newcommand{\ssc}       {\sin^{2} \chi}
\newcommand{\eom}       {\epsilon_{\Omega}}
\newcommand{\taugt}	{\tau_{g, \theta }}

\newcommand{\hmax}      {h_{\rm max}}

\newcommand{\hteetd}    {h(\taue = \taud)}
\newcommand{\ftheta}    {f_{\theta}}
\newcommand{\ub}        {u_{\rm break}}
\newcommand{\ubmt}      {\left(   \frac{u_{\rm break}}{10^{-3}}\right)}
\newcommand{\nd}	{\mbox{\boldmath$n_{\rm d}$}}
\newcommand{\ed}        {\epsilon_{\rm d}}
\newcommand{\eo}   	{\epsilon_0}

\newcommand{\dtsn}      {\Delta t_{\rm SN}}
\newcommand{\Dr}	{\Delta r}
\newcommand{\dtsnth}	{\left( \frac{\dtsn}{30 \rm \,yr} \right)}

\newcommand{\rdtsnth}	{\left( \frac{30 \rm \, yr}{\dtsn} \right)}
\newcommand{\rtauot}	{\left( \frac{10^{3} \rm \, yr}{\tau} \right)}
\newcommand{\Dtone}     {\Delta t_{1}}
\newcommand{\Dttwo}     {\Delta t_{2}}
\newcommand{\that}	{\hat{\theta}}

\newcommand{\si}{\sin i}
\newcommand{\ci}{\cos i}

\newcommand{\Icr}	{I_{\rm crust}}

\newcommand{\bmfive}	{\left(   \frac{b}{10^{-5}}   \right)}
\newcommand{\rr}	{\left( \frac{\rm kpc}{r} \right)}
\newcommand{\thmax}	{\theta_{\rm max}}
\newcommand{\ubtmt}	{\left( \frac{\ub}{10^{-3}} \right) }

\newcommand{\tauapd}      {\tau_{a, \phid}}
\newcommand{\tauat}      {\tau_{a, \theta}}

\newcommand{\nfour}	{\left(\frac{n}{10^{4}}\right)}
\newcommand{\nseven}	{\left(\frac{n}{10^{7}}\right)}
\newcommand{\Icrff}	{\left(\frac{\Icr}{10^{44} \rm \, g \, cm^{2}}\right)}

\title{Gravitational waves from freely precessing neutron stars}

\author[D. I. Jones and N. Andersson]
{D. I. Jones and N. Andersson\\
Faculty of Mathematical Studies, University of Southampton, 
Highfield, Southampton, SO17 1BJ, United Kingdom \\
}

\begin{document}

\maketitle

\begin{abstract}

In this paper we model the gravitational wave emission of a freely
precessing neutron star.  The aim is to estimate likely source strengths,
as a guide for gravitational wave astronomers searching for such signals.
We model the star as a partly elastic, partly fluid body with quadrupolar
deformations of its moment of inertia tensor.  The angular amplitude of the
free precession is limited by the finite breaking strain of the star's
crust.  The effect of internal dissipation on the star is important, with
the precession angle being rapidly damped in the case of a star with an
oblate deformation.  We then go on to study detailed scenarios where free
precession is created and/or maintained by some astrophysical mechanism.
We consider the effects of accretion torques, electromagnetic torques,
glitches and stellar encounters.  We find that the mechanisms considered
are either too weak to lead to a signal detectable by an Advanced LIGO
interferometer, or occur too infrequently to give a reasonable event rate.
We therefore conclude that, using our stellar model at least, free
precession is not a good candidate for detection by the forthcoming laser
interferometers.

\end{abstract}


\section{Introduction}

Freely precessing neutron stars have long been recognised as a potential
source of detectable gravitational waves \cite{zimm78,ap85}. Despite their
regular inclusion in review articles \cite{thor87,flan98} they have
received little in the way of detailed modelling.  In this paper we will
combine many of the physical processes relevant to the problem of free
precession and assess their relative importance.  The motivation behind
this work is to aid gravitational wave data analysis \cite{schu91}.  Given
the huge computational requirements of this analysis, any additional
information supplied by theoretical modelling of a source greatly increases
the chances of detection.  With the TAMA detector already operational, the
GEO600 and LIGO detectors due to go on-line within a year, and the VIRGO
detector following soon after, this issue is particularly pressing.

The study of gravitational wave generation from freely precessing neutron
stars can be divided into two parts.  The first problem is the formulation
of a free precession model consistent with our current understanding of
neutron star structure.  Neutron stars are not rigid bodies---they consist
of a thin elastic shell containing a superfluid core.  A model taking these
features into account was described in detail in Jones \& Andersson (2001),
where the effect of free precession on electromagnetic pulsar signals was
described and compared with pulsar observations.  (A handful of potential
free precession candidates were identified, but, as we will see, they all
rotate too slowly to be of gravitational wave interest).  A brief summary
of this model is set out in this paper.  A key feature is the decay of free
precession due to dissipative processes internal to the star.

The second part of the problem is to look at particular scenarios in which
free precession is created and/or maintained by an astrophysical mechanism.
The torques due to accretion disks, neutron star magnetic dipole moments,
and other gravitating bodies will be considered, as well as perturbations
associated with glitches.  We wish to investigate whether these mechanisms
are capable of balancing the internal dissipation to give steady long-lived
precessional motions.  Having done so, we will then be in a position to
estimate possible gravitational wave amplitudes in these various scenarios.

The structure of this paper is as follows.  In section \ref{sect:amonsfp}
we briefly describe our free precession model, and parameterise the
timescale in which internal dissipation damps the wobble motion.  In
section \ref{sect:tgwf} we show how the finite breaking strain of the crust
can be used to place an upper bound on the gravitational wave field of a
precessing star, regardless of its environment.  We also suggest a
detection strategy for gravitational wave data analysts.  General formulae
relating the gravitational wave amplitude to the strength and nature of a
pumping torque are given in section \ref{sect:copm}.  The possible
gravitational wave field strengths in various scenarios are then estimated
in sections \ref{sect:acctorq}--\ref{sect:cide}.  Our conclusions are given
in section \ref{sect:conc}, together with some suggestions for further
work.

\section{A model of neutron star free precession}
\label{sect:amonsfp}

\subsection{Dynamics of free precession}

In this section we will briefly summarise our model of neutron star free
precession.  For more detail see Jones \& Andersson (2001).  We will begin
by describing the free precession of a rigid body, as the motion in the
more realistic elastic shell/fluid core case can be thought of as a
modification of this.

The moment of inertia tensor of any axisymmetric rigid body can be written
as
\begin{equation}
{\bf I} =    I_0 \bdelta + \Delta I_{\rm d} ({\bf n_d n_d} 
         - \bdelta/3),
\end{equation}
where $\bdelta$ is the Kronecker delta, and the unit vector ${\bf n_d}$
points along the body's symmetry axis.  Then the principal moments are
$I_{1} = I_{2} = I_0 -\Delta I_{\rm d}/3$, $I_{3} = I_0 + 2\Delta I_{\rm
d}/3$, so that $I_{3}-I_{1} = \Delta I_{\rm d}$.  When $\DId >0$ the body
is said to be \emph{oblate}, and when negative the body is \emph{prolate}.
As we will describe below, the oblate case is the more physically
plausible.

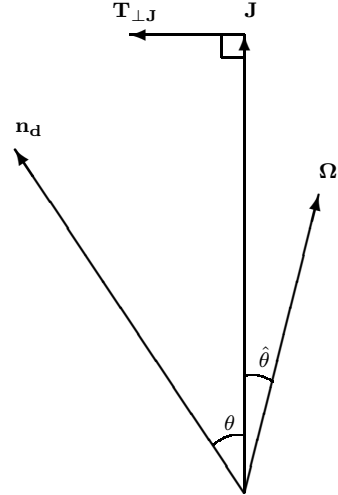
\begin{figure}
\begin{picture}(100,100)(-50,0)

\thicklines

\put(0,80){\vector(-1,0){20}}

\put(-23,83){$\bf T_{\perp J}$}

\put(0,76){\line(-1,0){4}}
\put(-4,76){\line(0,1){4}}

\put(0,0){\vector(0,1){80}}

\put(0,0){\vector(-2,3){40}}

\put(0,83){$\bf{J}$}

\put(0,0){\vector(1,4){13}}

\thinlines

\qbezier(-5.547,8.321)(-2.8,10.6)(0,10)


\qbezier(0,20.4)(2.5,21)(4.851,19.403)

\put(13,55){$\bf{\Omega}$}

\put(-40,63){$\bf{n_{d}}$}

\put(-3.5,11){$\bf{\theta}$}


\put(2.5,22){$\hat \theta$}

\end{picture}
\caption{This figure shows the reference plane for a freely precessing
body, which contains the deformation axis $ {\bf n_{d}}$, the angular
velocity vector ${\bf \Omega}$ and the fixed angular momentum ${\bf J}$.
The vectors ${\bf n_{d}}$ and ${\bf \Omega}$ rotate around $\bf{J}$ at the
\emph{inertial precession frequency} $\dot{\phi}$.  We refer to $\theta$ as
the \emph{wobble angle}.  The vector $\bf T_{\perp J}$ is the part of an
applied torque that causes a secular evolution in the wobble angle. }
\label{refplane}
\end{figure}

The angular momentum is related to the angular velocity according to
\begin{equation}
{\bf J} = (I_0 - \Delta I_{\rm d}/3) 
          {\bf \Omega} - \Delta I_{\rm d} \Omega_{3} {\bf n_d},
\end{equation}
where the 3-axis lies along ${\bf n_d}$.  This shows that the three vectors
${\bf J, \Omega}$ and ${\bf n_d}$ are always coplanar.  Following Pines \&
Shaham (1972a,b) we will call the plane so defined the \emph{reference plane}
(see figure \ref{refplane}).
Given that the angular momentum is fixed, this plane must revolve around
${\bf J}$.  The free precession is conveniently parameterised by the angle
$\theta$ between ${\bf n_{\rm d}}$ and {\bf J}.  We will refer to this as
the \emph{wobble angle}.  For a nearly spherical body the angle
$\hat{\theta}$ between ${\bf \Omega}$ and ${\bf J}$ is much smaller than
the angle between ${\bf J}$ and ${\bf n_d}$, according to
\begin{equation}
\label{eq:rigid_theta_hat}
\hat{\theta} \approx \frac{\Delta I_{\rm d}}{I_{1}} \sin \theta \cos \theta.
\end{equation}
We will denote by ${\bf n_{J}}$ the unit vector along
${\bf J}$.  Decomposing the angular velocity according to
\begin{equation}
{\bf \Omega} = \dot{\phi} {\bf n_{J}} + \dot{\psi} {\bf n_d}
\end{equation} 
then gives 
\begin{equation}
J = I_{1} \dot{\phi},
\end{equation}
\begin{equation}
\label{psidotrigid}
\dot{\psi} = - \frac{\Delta I_{\rm d}}{I_{3}} \dot{\phi}.
\end{equation}
The symmetry axis ${\bf n_d}$ performs a rotation about ${\bf J}$ in a
cone of half-angle $\theta$ at the angular frequency $\dot{\phi}$.  We will
refer to this as the \emph{inertial precession frequency}.  There is a
superimposed rotation about the symmetry axis ${\bf n_d}$ at the angular
velocity $\dot{\psi}$ .  This is usually referred to as the \emph{body
frame precessional frequency}, with the corresponding periodicity known as
the \emph{free precession period}:
\begin{equation}
P_{\rm fp} = \frac{2\pi}{\dot \psi}.
\end{equation}
For a nearly spherical body equation (\ref{psidotrigid}) shows that
$\dot{\psi} \ll \dot{\phi}$.  Note that the angles $(\theta, \phi, \psi)$
are simply the usual Euler angles which describe the orientation of the
rigid body (see e.g. Landau \& Lifshitz 1976, figure 47).

Turning now to the more realistic case of an elastic crust with a liquid
core, the moment of inertia of the crustal shell can be written as \cite{ap85}:
\begin{equation}
\label{eq:elasticmoi}
{\bf I} = I_0 \bdelta 
        + \Delta I_{\Omega} ({\bf n_{\Omega}n_{\Omega}} - \bdelta/3)
        + \Delta I_{d} ({\bf n_{d}n_{d}} - \bdelta/3).
\end{equation}
The first term on the right hand side is the moment of inertia of the
non-rotating undeformed spherical shell.  The second term is the change due
to centrifugal forces, and has ${\bf n_{\Omega}}$, the unit vector along
${\bf \Omega}$, as its symmetry axis.  The third term is the change due to
some other source of deformation, (such as strains in the crustal lattice),
and has the unit vector ${\bf n_{d}}$, fixed in the crust, as its symmetry
axis.  Without this third part the above equation would simply represent a
fluid ball, and free precession would not be possible.

The free precession of such an elastic shell containing a liquid core is
then very similar to the rigid result, with the geometry described in
figure 1 still applying, providing the $\Delta I_{\rm d}$ in equations
(\ref{eq:rigid_theta_hat}) and (\ref{psidotrigid}) is now set equal to the
deformation in the inertia of the \emph{whole} star, i.e. crust \emph{and}
core, while $I_{0}$ is still equal to the crustal moment of inertia only.
Explicitly:
\be
\label{realisticthat}
\that = \frac{\DId}{I_{\rm crust}} \theta,
\ee
\be
\label{realisticpsid}
\psid = -\frac{\DId}{I_{\rm crust}} \phid.
\ee
Note that centrifugal deformation $\Delta I_{\Omega}$ is not of importance
when considering the free precession geometry.

One further component can be added to our model: A pinned superfluid
coexisting with the inner crust.  As was first described by Shaham (1977),
the effect of such a component is to increase the body frame free
precession frequency $\psid$.  In the case where the rotation rates of the
superfluid and crust are the same, equations (\ref{realisticthat}) and
(\ref{realisticpsid}) still apply, with $\DId$ now containing a part equal
to the moment of inertia of the pinned superfluid.  However, as described
in Jones \& Andersson (2001), the handful of free precession candidates
identified in the pulsar population don't seem to have such a component.
Therefore, on those few occasions in this paper where we assume a
particular source of deformation, we will assume that $\DId$ is caused
entirely by Coulomb forces in the crustal lattice.  (If real stars do have
a significant pinned superfluid component, then the increased body frame
precession frequency will lead to an even faster dissipation of the free
precession energy, and tend to \emph{re-enforce} our final conclusion).

The effect of a torque on the free precession can be easily summarised.
Given that the torque-free motion is determined completely by the two
numbers $(\phid, \theta)$, we need only describe the effect of the torque
on these.  If the torque causes the magnitude of the angular momentum to
change at a rate $\Jdot$ then 
\be
\label{introapproxphidd}
\phidd = \frac{\Jdot}{I_{1}}. 
\ee
The evolution in the wobble angle is determined by the component of the
torque projected into the reference plane which lies perpendicular to
$\bfJ$:
\be
\label{introthetad}
\td =  -\frac{    T_{\perp J}} {\Icr \phid} , 
\ee
as  illustrated in figure \ref{refplane}.

\subsection{Sources of deformation}
\label{sect:sod}

The centrifugal deformation described above can be conveniently
parameterised by the dimensionless quantity $\eom$ which we will define by
$3\eom/2 = \DIOm / I_{\rm star}$ (this follows the notation of Alpar \&
Pines 1985).  This will be of the order of the rotational kinetic energy of
the star divided by its gravitational binding energy:
\begin{equation}
\label{epsilonfluid}
\eom \approx \frac{\Omega^{2} R^{3}}{GM}
      \approx 2.1 \times 10^{-3} \left(\frac{f}{100 \, \rm Hz}\right)^{2}
      R_6^3 / M_{1.4}
\end{equation}
where $\Omega =  2 \pi f$ is the angular frequency,  $R_6$ the neutron star
radius  in units  of $10^6$cm,  and  $M_{1.4}$ the  mass in  units of  $1.4
M_\odot$.

The deformation $\DId$ we can similarly parametrise in dimensionless form
using the relation $3\ed/2 = \DId / I_{\rm star}$.  This deformation is due
to some physical process other than rotation, which need not be specified
in many of our gravitational wave estimates.  However, in practice the most
significant source of deformation in a neutron star is likely to be strains
in its solid crust.  As described in Baym \& Pines (1971) and Pines \&
Shaham (1972a,b), if the crust has a zero strain oblateness $\eo$, the
actual deformation produced is of order
\begin{equation}
\label{DId}
\ed = \frac{B}{A+B} \epsilon_0.
\end{equation} 
where the constant $A$ depends on the stellar equation of state, and will
be of the order of the gravitational binding energy of the star.  The
constant $B$ also depends on the equation of state, and will be of order of
the total electrostatic binding energy of the ionic crustal lattice.  We
will define $b = B/(A+B)$ as the \emph{rigidity parameter}.  It is equal to
zero for a fluid star ($B=0$) and unity for a perfectly rigid one ($B/A
\rightarrow \infty$).  Realistic neutron star equations of state imply that
$b$ takes a value of:
\begin{equation}
\label{bestimate}
b \approx 1.6 \times 10^{-5} R_6^5 / M_{1.4}^3.
\end{equation}
(See Jones (2000) for a simple derivation, and Ushomirsky, Cutler \&
Bildsten (2000) for a detailed numerical treatment). The smallness of this
number reflects the fact that gravitational forces dominate crustal Coulomb
forces in determining the equilibrium shape of the star.  In this sense,
neutron star crusts are very far from perfectly rigid.  Note that $b \sim
10^{-5}$ for a canonical $1.4 M_{\odot}$, $10$ km neutron star.  More rigid
stars can exist only if less massive (and therefore larger radii)
neutron stars occur in nature, \emph{or} if current equations of state
seriously underestimate the crust thickness.  We will therefore present
results for rigidity parameters over the interval $10^{-3} \rightarrow
10^{-5}$, but will bear in mind that values at the smaller end of the
interval are more plausible.

For a very young star the zero-strain oblateness $\eo$ will be determined
simply by the star's shape at the moment when its crust first solidified,
i.e. $\eo = \eom(f=f_{\rm solid})$, where $f_{\rm solid}$ is the spin
frequency at the moment of solidification.  For older stars, $\eo$ will
have changed due to plastic deformation in the crust, either because of a
gradual creep or a more violent shape change, possibly connected with a
glitch.  Certainly, it seems likely that real stars will be oblate ($\eo >0
\Rightarrow \DId>0)$ rather than prolate ($\eo < 0 \Rightarrow \DId <0$).

\subsection{Wobble damping}
\label{sect:wd}

A real neutron star, once set into free precession, will not precess
forever---energy will be dissipated within the star, converting the kinetic
energy of the wobble into thermal energy.  Also, gravitational wave energy
and angular momentum will be radiated to infinity, which must be subtracted
off the star's motion.  

The problem of gravitational radiation reaction was examined in detail by
Cutler \& Jones (2000), who showed that the main result was to cause the
wobble angle to decay exponentially on a timescale:
\[
\tau_{\theta} = 1.8 \times 10^9 {\rm \, yr \,} \left(\frac{I_{\rm crust}}
{10^{44}\, {\rm g \, cm^{2}}}\right )
\]
\be
\label{finaltd}
\hspace{3cm}
\left( \frac{10^{38} \, {\rm g \,
cm^{2}}}{\Delta I_d}\right)^2 \, \left(\frac{100 \rm Hz}{f}\right)^4,
\ee
regardless of whether the deformation is oblate or prolate.

This is almost certainly much longer than the timescales connected with
internal dissipation.  In particular, models of neutron star interiors,
motivated in part by the need to model glitches, predict a frictional type
coupling between the crust and core. The result of this coupling is a torque
${\bf T}$ exerted on the crust:
\begin{equation}
{\bf T} = K({\bf \Omega_{fluid} - \Omega_{solid}}),
\end{equation}
where $K$ is a positive constant.  Such a torque would tend to restore
corotation between crust and core in a glitching neutron star.  However, as
modelled by Bondi \& Gold (1955), the torque would also tend to damp the
wobble angle of a precessing body.  The timescale for this damping can be
parameterised by $n$, the number of free precession periods $P_{\rm fp}$ in
which one e-fold occurs: 
\begin{equation} 
\taud = \frac{I_0}{\DId} n P.  
\end{equation} 
where $P$ is the spin frequency (approximately $2 \pi /\psid$) of the body.
The parameter $n$ has been estimated by Alpar \& Sauls (1988) who examined
the scattering of electrons off the superfluid vortices.  This interaction
is sometimes referred to as `mutual friction'.  Alpar \& Sauls estimated
$n$ to lie in the interval $400 \rightarrow 10^4$, giving a wobble damping
timescale of:
\[
\taud  = 3.2 \, {\rm yr} \, 
          \left(\frac{n}{10^4}\right)
          \left(\frac{100 \rm \, Hz}{f}\right)
\]
\be
\label{introtaud}
\hspace{3cm}
          \left(\frac{I_0}{10^{44} \, \rm g \, cm^2}\right)
          \left(\frac{10^{38} \, \rm g \, cm^2}{I_{\rm d}}\right).
\ee
Comparing equations (\ref{finaltd}) and (\ref{introtaud}) it is clear that
internal dissipation is likely to damp the free precession much more
rapidly that gravitational radiation reaction.  

Interestingly, in the case of \emph{prolate} deformations, the internal
dissipation acts so as to \emph{increase} the wobble angle.  However, as
described in section \ref{sect:sod}, prolate deformations are probably not
likely to be found in real stars, and so for the remainder of this paper we
will assume that deformations are oblate, corresponding to a damping of
the wobble motion.

\section{The gravitational wave field}
\label{sect:tgwf}

\subsection{General form of the field}

The gravitational wave field for a rigid precessing body was first
calculated by Zimmerman \& Szedenits Jr (1979), using the mass quadrupole
formalism (see e.g. Misner, Thorne and Wheeler 1973).  As described in the
previous section, even though the elasticity and fluid core are important
in determining the free precession period, the geometry of free precession
in the realistic case is very similar to that in the rigid case---the
deformation bulge moves in a cone of half-angle $\theta$ at a rate $\dot
\phi$.  (The superimposed rotation at $\dot \psi$ about the axis $\nd$ does
not change the mass quadrupole of the body, and so does not appear in the
gravitational wave physics).  It follows that, to the accuracy of our model
at least, the wave field calculated by Zimmerman \& Szedenits Jr (1979)
applies in the realistic case.  Explicitly, the gravitational waves are
emitted at frequencies $\phid$ and $2\phid$ and with the two polarisations
(denoted by $+$ and $\times$):
\be
\label{hpone}
h_{+} (\phid)       =  \frac{2\phid^{2}}{r}  \si \ci 
                      \DId \st \ct  \cos (\phid t) 
\ee
\be
\label{hptwo}
h_{\times} (\phid)  =  \frac{2\phid^{2}}{r}  \si  
                      \DId \st \ct \sin (\phid t) 
\ee
\be
\label{hcone}
h_{+} (2\phid)      =  \frac{2\phid^{2}}{r}  (1 + \cos^{2} i)
                      \DId \sst  \cos (2\phid t)
\ee
\be
\label{hctwo}
h_{\times} (2\phid) = \frac{2\phid^{2}}{r} 2 \ci \DId \sst \sin (2\phid t)
\ee
for a source at distance $r$ and angular momentum at inclination angle
$i$ to the line of sight. Note that the $\DId$ factor is the part of the
quadrupole moment tensor due to the deformation process, \emph{not} the
centrifugal piece.  The modifications to this result due to the non-rigidity
are small, and lie beyond the accuracy of our free precession model.

\subsection{Limit on gravitational wave amplitudes due to finite crust
strength}

As described in Jones \& Andersson (2001), a real neutron star crust will
have a finite breaking strain $u_{\rm break}$.  This can be used to place
an upper bound on the wobble angle of a precessing star.  We know that (for
small wobble angles at least) when an initially axially symmetric body is
set into free precession, a deformation $\Delta I_{\rm d}$ remains along the
axis ${\bf n_{d}}$ fixed in the star, while a deformation $\Delta
I_{\Omega}$ points along the angular velocity vector.  From the point of
view of an observer attached to the crust, a deformation of size $\Delta
I_{\Omega}$ describes a cone of half-angle $\theta + \hat{\theta} \approx
\theta$ about ${\bf n_d}$.  This change in shape is all we need to know to
estimate the strain: The change in position of any given particle is of
order $R \epsilon_{\Omega} \theta$, while the corresponding strain is of
order $\epsilon_{\Omega} \theta$.  This precession-induced strain is not
constant, but varies with magnitude $\epsilon_{\Omega} \theta$ over one
(body frame) free precession period.  As there exists a maximum strain
$u_{\rm break}$ that the solid can withstand prior to fracture, the wobble
angle will be limited to a value of $u_{\rm break}/\epsilon_{\Omega}$ so
that: 
\begin{equation}
\label{eq:thetamax}
\theta_{\rm max} \approx 0.45 \left(\frac{100 \, \rm Hz}{f}\right)^{2} 
                 \left(\frac{u_{\rm break}}{10^{-3}}\right) {\rm \, radians}.
\end{equation}
Qualitatively, we can say that faster spinning neutron stars have larger
bulges to re-orientate and therefore can sustain smaller wobble angles
prior to fracture.  For sufficiently slowly spinning stars the above
equation breaks down, yielding angles in excess of $\pi/2$.  The wobble
angles of such slowly spinning stars are not limited by crustal strain.

The value of $u_{\rm break}$ is highly uncertain.  By extrapolating the
breaking strains of terrestrial materials, Ruderman (1992) suggests the
value relevant for neutron star crusts may lie in the range $10^{-2}$ to
$10^{-4}$.

This limit on the wobble angle $\theta$ can be used to place an upper bound
on the gravitational wave amplitude of a freely precessing star with a
given deformation $\DId$.  To do so we will characterise the field strength
by
\be
\label{hamp}
h = \frac{G}{c^{4}} \frac{\phid^{2}}{r} \DId \theta.  
\ee 
This is an order-of-magnitude approximation to the set of equations
(\ref{hpone})--(\ref{hctwo}).  The trigonometric factor describing the
variation of wave amplitude with inclination angle $i$ has been neglected
entirely, while the trigonometric factor in the precession angle $\theta$
has been replaced by its small angle limit.  Note that in this small-angle
limit the radiation is emitted mainly at the frequency $\phid$---the
radiation at frequency $2\phid$ is a factor $\theta$ smaller.

The purpose of this approximation is to enable us to gain a insight as to
how gravitational wave amplitudes depend upon source parameters such as the
breaking strain, frequency and deformation size.  This analysis is a useful
preparation for sections \ref{sect:acctorq}--\ref{sect:cide} where detailed
astrophysical situations are considered.

We will specialise to the case where the deformation $\DId$ is due to
Coulomb forces in the crustal lattice.  We set $\DId = 3I_{\rm star} \ed
/2$ with $\ed$ given by equation (\ref{DId}).  For definiteness, we will
assume the crust is `relaxed', i.e. the zero-strain oblateness $\eo$ is
equal to the centrifugal oblateness $\eom$.  (These two quantities can
differ by the small crustal breaking strain anyway---see Jones \& Andersson
2001).  We then find
\be
\label{hcpar}
h = 3.7 \times 10^{-28} \foh^{4} \bmfive \rr \theta
\ee
for a spin frequency $f$.  In (\ref{hcpar}) $\theta$ has been left a free
parameter.  However, to obtain an upper bound on $h$ we can set $\theta$ to
its maximum value as obtained from crust cracking considerations, i.e. put
$\theta = \thmax$.  We then obtain
\be
h (\theta = \thmax) = 3.7 \times 10^{-28} \foh^{4} \bmfive
\ee
for $f < \ftheta$ and
\be
h (\theta = \thmax) = 1.8 \times 10^{-28} \foh^{2} \bmfive \ubtmt
\ee
for $f > \ftheta$, where $\ftheta$ is the frequency at which $\thmax = 1$:
\be
\ftheta = 69 {\rm \, Hz} \ubtmt^{1/2}
\ee
For spin frequencies close to and below this our small-angle approximation
breaks down, so we have put $\theta = 1$ for $f < \ftheta$ so as to still
obtain results correct to within an order-of-magnitude.

\begin{figure}
   \centerline{ \psfig{file=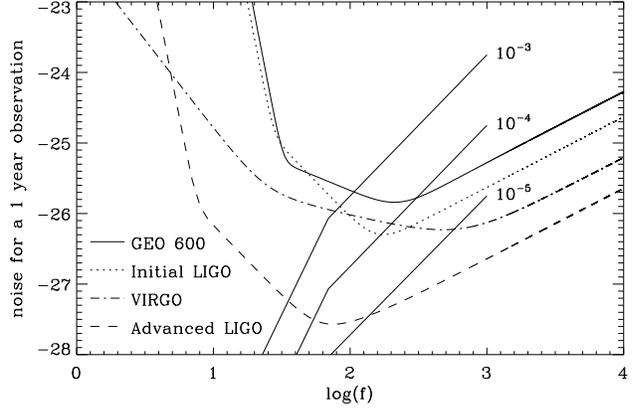,width=9cm} } \caption{The maximum
      gravitational wave amplitude for Coulomb deformations.  The star is
      at a distance of 1 kpc with $\ub = 10^{-3}$ and $b = 10^{-3},
      10^{-4}, 10^{-5}$.  The matched filter has been assumed to accumulate
      signal for an interval of one year.
      The noisecurves have been taken from Owen \& Sathyaprakash (1999).}
      \label{csone}
\end{figure}

\begin{figure}
  \centerline{
      \psfig{file=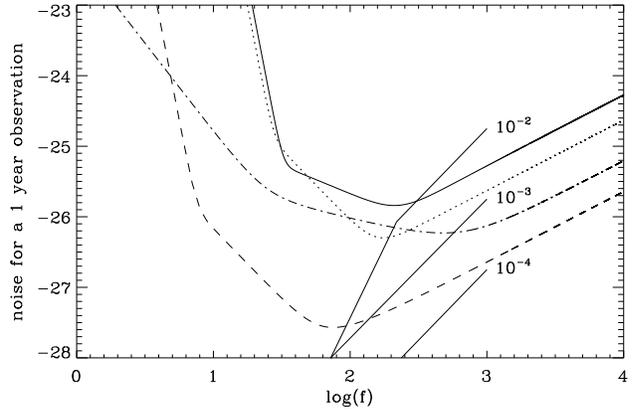,width=9cm}   }
   \caption{The maximum gravitational wave amplitude for Coulomb deformations.  The star is at a distance of 1 kpc with $b = 10^{-5}$ and $\ub = 10^{-2}, 10^{-3}, 10^{-4}$.  The noise curves are as indicated in figure \ref{csone}. }
   \label{cstwo}
\end{figure}

In figures \ref{csone} and \ref{cstwo} we have plotted the maximum
amplitude $h$ for a variety of neutron star parameters.  The noisecurves
have been taken from Owen \& Sathyaprakash (1999).  Note that a knee
appears in many of the signal curves.  This knee corresponds to $f =
\ftheta$.  Above this frequency the wobble angle is limited according to
(\ref{eq:thetamax}).  We have assumed the matched filtering can accumulate
signal only for a time of one year.  The amplitudes are shown only for
frequencies less than a kilohertz, as stars rotating more rapidly than this
have not yet been observed.  Figure \ref{csone} plots $h$ for the
parameters $\ub = 10^{-3}$, $r = 1$ kpc and for three different values of
$b$: $b=10^{-5}$, $b=10^{-4}$, $b=10^{-3}$.  Recall that $b=10^{-5}$ is the
value expected for a canonical $1.4 M_{\odot}$ $R=10$ km neutron star,
while $b=10^{-3}$ would correspond to a lighter larger star with a thicker
crust.
In figure \ref{cstwo} plots $h$ for $b= 10^{-5}$ and breaking strains of
$\ub= 10^{-2}, 10^{-3}$ and $10^{-4}$.

In making the above assumptions regarding matched filtering we have assumed
that the wobble angle and spin frequency of the star remain constant for
the observation period of one year.  In practice the wobble angle will
decay significantly over this interval (see section \ref{sect:wd}) unless a
sufficiently strong pumping mechanism is active.  However, the purpose of
these figures isn't to model in detail any particular scenario---they are
included simply so that we might gain insight into how wave amplitudes
depend upon the rigidity and breaking strain parameters, and identify
parameters values capable of leading to detectable signals.  These figures
say nothing about the distance to the nearest source, or, in the case of
burst-like sources, the event rate.  Issues such as these can only be
addressed in the case of particular pumping mechanisms.  This will be
carried out in sections \ref{sect:acctorq}--\ref{sect:cide}, where the
complicating effect of internal dissipation will be included also.  In
short, figures \ref{csone} and \ref{cstwo} represent absolute upper bounds
on the wave amplitude for a given spin frequency, crustal breaking strain
and deformation, but give no information about likely wave amplitudes in
nature.

With these qualifications in mind, we will note that these figures show
that for neutron stars with $b \ga 10^{-3}$, the $\theta = \thmax$
gravitational wave signal is detectable by first-generation interferometers
for frequencies in excess of 100 Hz.  Slower spinning sources
produce steady gravitational wave signals which are intrinsically too weak
to be detected.  Stars with $b \approx 10^{-4}$ are potentially detectable
by first generation interferometers only for frequencies close a 1 kHz.
More realistically, stars with $b \approx 10^{-5}$ are potentially
detectable by an Advanced LIGO interferometer at frequencies over 100 Hz.

\subsection{Detection strategy}

The problem of determining source parameters from electromagnetic
observations of a freely precessing neutron star was examined by Jones \&
Andersson (2001).  The extraction of these parameters in the case of a
gravitational wave detection was examined by Zimmermann \& Szedenits
Jr. (1979), who showed that measurement of all four components of $h$ given
in equations (\ref{hpone})--(\ref{hctwo}) allows extraction of the
inclination angle $i$, the wobble angle $\theta$ and the quantity $\DId
/r$.

As was also noted by Zimmermann \& Szedenits Jr. (1979), if this is
combined with an estimate of $r$ a value for $\DId$ is obtained.  This is
of considerable use: If the deformation is of Coulomb origin the equation
$\DId = 3 I_{\rm star} b \eo/2$ can be combined with equation-of-state
calculations of the rigidity parameter $b$ to give the reference oblateness
$\eo$.  A value of $\eo$ very different from the fluid oblateness might
then be a sign that the deformation $\DId$ is \emph{not} due to crustal
strains.

Note however that if the wobble angle is very small the above two equations
can be solved only for $i$ and the product $\DId \theta /r$.  This is
likely to be the case for the fastest spinning stars whose wobble angles
are limited by fracture according to equation (\ref{eq:thetamax}).

If the source is observed as a pulsar it may be possible to obtain
additional information.  Suppose the electromagnetic pulse originates from
a dipole $\bf{m}$ inclined at an angle $\chi$ to the deformation axis.
Then the motion of $\bf{m}$ is due to the combined rotation of $\nd$ about
$\bf{J}$ at $\phid$ and the motion of $\bf{m}$ about $\nd$ at $\psid$.  As
noted by Zimmermann \& Szedenits Jr. (1979) the electromagnetic pulse
frequency is then given by $ f_{\rm em} = \phid + \psid$ for $\theta <
\chi$, while for $\theta > \chi$ we have $ f_{\rm em} = \phid$.

In the case of Coulomb deformations of rapidly spinning stars we would
expect the $\theta < \chi$ regime to apply.  Then the difference between
electromagnetic and gravitational wave frequencies would give $\psid$, and
to leading order equation (\ref{realisticpsid}) gives
\be 
\DId = \Icr \psid / f_{\rm em}.
\ee 
If in addition an estimate of the pulsar's distance were available the
product $(\DId/\Icr) \theta /r$ could be decomposed into its component
parts.

The above remarks concerning frequency splitting suggest the following
strategy for gravitational wave observers.  Suppose we are searching for a
precessional gravitational wave signal from a star of known (average)
electromagnetic pulse frequency $f$.  Then one matched filter should be
used at that frequency to cover the $\theta > \chi$ case where there is no
splitting between the gravitational and electromagnetic signals.  However,
if splitting does occur, it will be necessary to search at a rotational
frequency $\Delta f = f\DId/\Icr$ above this.  For definiteness we will
assume a Coulomb deformation to give
\be
\label{deltaf}
\Delta f = 3 \times 10^{-5} \foh^{3} \bmfive \left(\frac{I/\Icr}{10}\right)
\ee
where equation (\ref{epsilonfluid}) has been used.  

We now wish to identify a frequency band in which to search for the
gravitational signal.  This is easily done.  Consider some particular
frequency, say $100$ Hz.  Then from figure (\ref{csone}) we see that even
for an Advanced LIGO, $b$ values less than $10^{-5}$ give signals too weak
to be detected.  From the above formula this value of $b$ corresponds to
$\Delta f = 3 \times 10^{-5}$ Hz.  Therefore there is no point in searching
for signals at frequencies less than ($100 + 3 \times 10^{-5}$) Hz (apart
from the signal at $f$ mentioned above).  On the other hand, on physical
grounds the value $b = 10^{-3}$ is surely an upper bound on the rigidity
parameter.  This corresponds to $\Delta f = 3 \times 10^{-3}$ Hz.
Therefore there is no point in searching for signals at frequencies greater
than ($100 + 3 \times 10^{-3}$) Hz.  In this way we identify a frequency
band in which to fruitfully search for the gravitational signal.  A
year-long integration would require matched filters with a $3 \times
10^{-8}$ Hz spacing.  In this example this would require approximately
$10^{5}$ templates.

The above argument attempts to identify the gravitational waves at the
inertial precession frequency $\phid$.  There will be radiation at $2
\phid$ also.  This can be searched for using templates at twice the
frequencies described above.

\section{Classification of pumping mechanisms}
\label{sect:copm}

It is useful to categorise pumping mechanisms according to the sort of
evolution they produce in the wobble angle.  As discussed above, the effect
of a torque depends upon its projection into the reference plane, which
rotates around the star's angular momentum vector at a rate $\phid$.  This
projection will depend upon the details of the mechanism producing the
torque, possibly leading to a complicated evolution in the wobble angle.
Nevertheless, we will find it useful to define the four categories
described below.  We will define $\taup$ as a timescale characterising the
strength of the pumping torque.  Its exact meaning will depend upon the
nature of the mechanism.

It is also useful to derive the following expressions for the wave
amplitude written in terms of a timescale rather than a frequency.
Equation (\ref{hamp}) can be combined with equation
(\ref{finaltd}) to give the wave amplitude in terms of the
gravitational alignment timescale: 
\be
\label{introhtaugt}
h = \left[\frac{5G}{2c^{3}} \frac{\Icr}{\taugt}\right]^{1/2} \frac{\theta}{r}.
\ee
This will be extremely useful when estimating wave amplitudes for
particular pumping mechanisms.  Combining equation (\ref{introhtaugt}) with
equation (\ref{eq:thetamax}) then gives an estimate of the maximum possible
wave amplitude at a given frequency and gravitational alignment timescale
$\tau_{g,\theta}$: 
\[
\hmax = 1.4 \times 10^{-24}   \Icrff^{1/2}
\]
\be
\label{introhmtgone}
\hspace{4cm} \left(\frac{10^{3} \rm \, yr}{\taugt}\right)^{1/2} \ronekpc
\ee
for $f<\ftheta$, and
\[
\hmax = 6.7 \times 10^{-25} \Icrff^{1/2} \ubmt
\]
\be
\label{introhmtgtwo}
\hspace{2cm}
\rfoh^{2}
            \left( \frac{10^{3} \rm \, yr}{\taugt} \right)^{1/2} \ronekpc
\ee
for $f>\ftheta$.

In a similar way equations (\ref{hamp}) and (\ref{introtaud}) can be
combined to give the wave amplitude in terms of the internal dissipation
timescale $\taud$:
\be
\label{introhtaud}
h = \frac{2\pi G}{c^{4}} \Icr  \frac{n \phid}{\taud} \frac{\theta}{r}.
\ee
Setting $\theta$ equal to its maximum value then gives
\[
\hmax = 3.3 \times 10^{-30} \nfour \foh 
\]
\be
\label{introhmtdone}
\hspace{2cm}
\Icrff  
        \left( \frac{10^{3} \rm \, yr}{\taud} \right) \ronekpc
\ee
for $f<\ftheta$, and
\[
\hmax = 1.6 \times 10^{-30} \nfour \ubmt \rfoh 
\]
\be
\label{introhmtdtwo}
\hspace{2cm}
\Icrff 
            \left( \frac{10^{3} \rm \, yr}{\taud} \right) \ronekpc
\ee
for $f>\ftheta$.  Again, these formulae will be of use in estimating wave
amplitudes in later sections.

\subsection{Oscillatory pumping}

We will first consider the case where the torque does not remain fixed
relative to the reference plane, but instead rotates around it at some
frequency.  In sections \ref{sect:acctorq} and \ref{sect:emtorq} we will
identify such torques that rotate at the body frame precessional frequency
with respect to the reference plane, i.e. with a frequency $\psid$.
Specialising to this case we see that for half a precession period the
wobble angle will increase, but for the next half-cycle it will decrease.
We will refer to this as \emph{oscillatory pumping}.  It will produce a
wobble angle that varies as
\be
\label{tosct}
\theta = \theta_{osc} (1 + \cos \psid t)
\ee
where $\theta_{osc}$ denotes the average value of $\theta$ attained.  To
order-of-magnitude accuracy this will be given by
\be
\label{esttosc}
\theta_{osc} \sim   \frac{\Pfp}{\taup} \sim \frac{\Icr}{\DId} 
                                            \frac{1}{\phid \taup}
\ee
where equation (\ref{realisticpsid}) has been used.  The wave field is then
estimated by substituting (\ref{esttosc}) into (\ref{hamp}) to give
\be
\label{hosc}
h \sim \frac{2 \pi G}{c^{4}}  \frac{\Icr}{r} \phid \frac{1}{\taup}.
\ee
Note that this is independent of the size of the deformation $\DId$.
However, in the case of sufficiently small $\DId$ the precession angle as
given by (\ref{esttosc}) will exceed $\thetam$.  In such a case equation
(\ref{hosc}) represents an upper bound.

\subsection{Exponential pumping}
\label{sect:genexp}

In subsequent sections we will find examples of torques which tend to
increase the wobble angle at a rate proportional to the angle itself,
i.e. produce \emph{exponential pumping}.  Then
\begin{equation}
\label{tdexp}
\thetad = \theta \left[ \frac{1}{\taup} - \frac{1}{\taugt} -
\frac{1}{\taud} \right].  
\end{equation}
In general, these timescales are all functions of frequency.  In the case
where the frequency and therefore the timescales are not constant the
solution will be complex.  However, in a constant frequency system such as
an accreting star at spin equilibrium, the above equation leads to
exponential solutions.  We will consider such a system.
   
Clearly, in order to calculate the wave amplitude due to such a mechanism
we need to compare the three timescales that appear in (\ref{tdexp}).  In
order to simplify this problem we will divide it into two smaller ones.
First we will include only the pumping and gravitational radiation
reaction.  This will lead to an upper bound on $h$ for stars free of
internal dissipation.  Then we will include the pumping and internal
dissipation only.  This will lead to an upper bound on $h$ for
gravitational radiation reaction free stars.  Real stars will be acted upon
by both gravitational radiation reaction and internal dissipation.
Therefore, at a given frequency, the upper bound on $h$ for real stars will
be \emph{less} than the \emph{minimum} of the two separate bounds.  In
fact, we have already seen that in practice internal dissipation is more
effective than gravitational radiation reaction in damping the wobble
(section \ref{sect:wd}).  We will nevertheless include gravitational
radiation reaction in this section, partly to quantify just how much weaker
it is than internal damping, and partly to be able to give an upper bound
on the gravitational wave amplitude that would apply even if the estimates
of internal damping strength of section \ref{sect:wd} are too large by many
orders of magnitude.

We begin by neglecting the internal dissipation.  From equation
(\ref{tdexp}) we see that $\theta$ evolves exponentially, increasing when
$\taugt > \taup$ and decreasing when the inequality is reversed.  The
wobble angle will then either increase to its maximum value or decrease to
zero:
\bea
\label{texpone}
\theta & = & \thetam \hspace{1cm} {\rm for} \hspace{1cm} \taugt > \taup \\
\label{texptwo}
\theta & = & 0    \hspace{1.5cm} {\rm for} \hspace{1cm} \taugt < \taup.
\eea
  
Substitution of (\ref{texpone}) into (\ref{introhtaugt}) then gives an
estimate of the wave amplitude:
\be
h = \left[\frac{5G}{2c^{3}} \frac{\Icr}{\taugt}\right]^{1/2} \frac{\thetam}{r}.
\ee
This is valid for $\taup < \taugt$ and so we obtain an \emph{upper bound} on $h$ if we set these two timescales equal:
\be
\label{hmaxexpg}
h(\taup = \taugt) =  
   \left[\frac{5G}{2c^{3}} \frac{\Icr}{\taup}\right]^{1/2} \frac{\thetam}{r}.
\ee

Now consider the case where the gravitational radiation is neglected.  Then
equations of the form of (\ref{texpone}) and (\ref{texptwo}) apply again,
but with $\taugt$ replaced by $\taud$.  Substitution into
(\ref{introhtaud}) then gives the estimate
\be
h = \frac{2 \pi G}{c^{4}} \frac{\Icr}{r} \phid n \frac{\thetam}{\taud}.
\ee
This is valid when $\taup < \taud$ and so an upper bound is obtained when
the timescales are set equal:
\be
\label{hmaxexpd}
h(\taup = \taud) = \frac{2 \pi G}{c^{4}} \frac{\Icr}{r} \phid n
                   \frac{\thetam}{\taup}.
\ee

\subsection{Linear pumping}

Now consider the case of a torque that is fixed with respect to the
reference plane.  Such a torque would lead to a linear variation in the
wobble angle, i.e. \emph{linear pumping}.  In this case
\be
\label{tdlin}
\thetad =  \frac{1}{\taup} 
         - \theta \left[\frac{1}{\taugt} - \frac{1}{\taud}\right].
\ee
We will break this up into two separate problems as above.

When the internal dissipation is neglected we see that there is an
equilibrium precession angle at which the effects of the two torques
balance:
\be
\label{tling}
\theta = \frac{\taugt}{\taup}.
\ee
Substitution into equation (\ref{introhtaugt}) then gives the following
estimate for the wave field:
\be
\label{hling}
h =  \left[\frac{5G}{2c^{3}} \Icr \right]^{1/2} \frac{1}{r} 
           \frac{\taugt^{1/2}}{\taup}.
\ee
Strictly this is an upper bound, as when $\taup$ is decreased below
$\taugt$ the precession angle as given in (\ref{tling}) becomes
unphysically large.  More precisely, the above equation represents an
estimate of $h$ for systems with $\taugt / \taup < \thetam$ but an upper
bound for systems where the inequality is reversed.

The analogous calculation for radiation reaction free systems gives
\be
\label{tlind}
\theta = \frac{\taud}{\taup}
\ee
leading to the upper bound
\be
\label{hlind}
h(\taup = \taud) = \frac{2 \pi G}{c^{4}} \frac{\Icr}{r} \phid
                   \frac{n}{\taup}.
\ee

Clearly the exponential and linear pumping mechanisms are very different.
The exponential mechanism leads to precession only when the pumping torques
overcome the radiation reaction and internal dissipation effects,
i.e. either gives wobble angles limited by $\thetam$ or gives no precession
at all.  On the other hand the linear pumping mechanism will always give a
finite precession angle, as indicated in equations (\ref{tling}) and
(\ref{tlind}).  Also, exponential pumping mechanisms cannot lead to
precession in systems which are originally not precessing, while linear
pumping mechanisms can.  However, the upper bounds on $h$ obtained via the
two types of pumping are similar.  Comparing (\ref{hmaxexpg}) with
(\ref{hling}) and (\ref{hmaxexpd}) with (\ref{hlind}) we see that the upper
bounds differ only by a factor $\thetam$ which is of order to unity for low
frequencies.

\subsection{Impulsive pumping}

We will describe any process that acts to increase $\theta$ on timescales
much less than a precession period as an \emph{impulsive pumping}
mechanism.  For an isolated star the angular momentum during this interval
will be constant (apart from any negligible amount carried away by a
radiation field, unless the star ejects part of its mass!).  It therefore
follows that the kinetic energy of the star must change.  On the other
hand, if the impulse is brought about via an interaction with another body
it is possible for only the kinetic energy to change, or only the angular
momentum, or both.  Glitches in young pulsars are of interest as candidates
for isolated impulsive pumping (section \ref{sect:natgl}), while near-body
encounters in dense environments are of interest as candidates for
non-isolated impulsive pumping (section \ref{sect:cide}).

\section{Accretion torques}
\label{sect:acctorq}

Accretion torques are an obvious place to start when looking for mechanisms
to pump precession.  Not only are they capable of exciting wobble, but they
can also maintain the spin frequency of the system, leading to the
possibility of long-lived constant wave amplitude sources.  Indeed,
accretion has already been investigated as a means of powering
gravitational wave emission via CFS-type instabilities \cite{pp78,wago84}
and via quadrupole moment asymmetries connected with crustal composition
variations (Bildsten 1998).

The torque on the central star is the sum of two parts.  The first is
simply the material torque, i.e. due to the accretion of angular momentum
from matter detaching from the disk and falling onto the star.  The second
torque is due to the coupling of the star's magnetic field with the disk.
The net effect of these torques is to spin-up slowly rotating stars but
impose a maximum spin frequency for fast stars where the accretion torque
vanishes. It is useful to construct an order-of-magnitude accretion spin-up
timescale by considering the material torque at the magnetosphere radius:
\be
\label{acctimescale}
\tauapd \approx 1.34 \times 10^{4} {\rm \, yr \,}  \fHz
        \left[ \Bnine \Medd^{3} \right]^{-2/7}
\ee
where $\dot{M}$ and $M_{\rm E}$ denote the actual and Eddington accretion
rates \cite{st83}.  Note, however, that this will only be equal to the
spin-up timescale for slowly rotating stars.  The accretion torque will
vanish when the corotation radius lies just outside the inner disk edge.
This limits the spin frequency of fast stars:
\be
\label{faccmax}
f_{\rm max} \approx 526 {\rm \, Hz \,}  \Bnine^{-6/7} \Medd^{3/7}.
\ee
This can be combined with equation (\ref{acctimescale}) to give a lower
bound on $\tauapd$ for a given frequency and accretion rate:
\be
\tauapd \ge 1.7 \times 10^{3} {\rm \, yr \,} \fHz^{4/3} \Medd.
\ee
If we assume that this torque has a significant component perpendicular to
the star's angular momentum vector, then the corresponding wobble pumping
timescale must be $\Icr/I$ times this (see equation \ref{introthetad}):
\be
\tauat \ge 1.7 \times 10^{3} {\rm \, yr \,} \fHz^{4/3} \Medd \frac{\Icr}{I}.
\ee
We will make use of this when estimating maximum wave amplitudes below.
		
Almost all analyses of accretion torques have assumed axisymmetry, giving
rise to either purely spin-up or spin-down torques.  However, precession
pumping requires a torque with a component orthogonal to the spin axis.  It
follows that if we wish to find accretion torques capable of pumping
precession we need to identify situations in which the torque itself would
be non-constant.  Fortunately, we would expect this to be the case when
magnetospheric effects are taken into account.  As pointed out by Lamb et
al. (1975), \nocite{llps75} the accretion rate and therefore also the
torque depend upon the balance of gravitational, centrifugal and magnetic
forces.  The torque is therefore a function of the plasma angular velocity,
stellar angular velocity and stellar magnetic moment vectors.  However, if
the star is itself precessing the relative orientation of these vectors
will be modulated.  This in turn must lead to a modulation of the accretion
torque locked in phase with the precession.  It is precisely this sort of
modulation that we would expect to lead to a secular evolution
in the wobble angle.

Lamb et al. (1975) demonstrated that such a modulation can be effective in
exciting large amplitude precession of a rigid body, with an excitation
timescale of the order of the spin-up timescale, i.e. of the order of
$\tauapd$ in equation (\ref{acctimescale}).  For our more realistic stellar
model this would correspond to an excitation on a timescale $\Icr/I$ times
shorter.  A full description of the conditions under which such pumping can
occur would require a detailed understanding of the accretion process.
Such a description is still not available. 

To re-enforce the potential complexity of this problem, the time variation
of each vector involved in this problem is summarised in table
\ref{acctable}.  Note that the vector $\ndisk$ describing the plane of the
disk is not well-defined.  Far from the central star the disk is likely to
be planar.  However, Vietri \& Stella (1998) have argued that a non-aligned
dipole will tend to lift plasma out of the disk plane.  In the inner disk,
viscosity will not be effective in preventing this out-of-plane motion.
The inner disk will then undergo a \emph{forced} precession with its normal
moving in a cone about the star's angular momentum.

A complete description of how this modulates the torque on the star has not
been attempted previously, and lies beyond the scope of this paper.
Nevertheless, it seems likely that at least part of the torque will be
locked in phase with the disk precession.  This torque would be most
effective in exciting stellar free precession if the inner disk rotated at
a rate close to the stellar spin frequency, or---more accurately---at the
frequency $\phid$.  However, for all sensible parameters the Lense-Thirring
precession frequency of the inner disk is much less than the star's spin
frequency.  This will lead to a rapid averaging of the effects of the
torque on the wobble angle.  It therefore seems unlikely that the forced
precession of the inner disk plays a crucial role in wobble pumping.  We
will simply note that the notion of a single fixed disk orientation is
flawed.  The vector $\ndisk$ in table \ref{acctable} should be regarded
as an \emph{average} orientation, which will describe well the outer part
of the disk, but not necessarily the inner part.
\begin{table*}
\footnotesize
\begin{minipage}{150mm}
\caption{This table summarises the temporal behaviour of the vectors of importance in the accretion problem.  The final column gives their rotation rate as viewed from the inertial frame.  As discussed in the text, the disk, particularly in its inner regions, may not be planar and stationary, so that $\ndisk$ below should be regarded as an average disk orientation.  Also note that the motion of $\bfm$ is rather complicated---it moves on a cone of half-angle $\chi$ about $\nd$, while $\nd$ itself moves on a cone of half-angle $\theta$ about $\bfJ$.  These two rotations combine to give the motion of $\bfm$ indicated.  In this case the angular velocities are average values.}
\label{acctable}
\begin{tabular}{|l|l|l|l|} \hline
Vector & Description & Motion & Frequency \\ $\ndisk$ & Normal to disk &
None & 0 \\ $\bfJ$ & Angular momentum of star & None & 0 \\ $\bfOm$ &
Angular velocity of star & Cone of half-angle $3\ed \theta/2$ about $\bfJ$
& $\phid$ \\ $\nd$ & Axis of star's deformation & Cone of half-angle
$\theta$ about $\bfJ$ & $\phid$ \\ $\bfm$ & Dipole axis & Cone,
$|\chi-\theta|<$ half-angle $<|\chi + \theta |$, & $\phid + \psid$ for
$\theta < \chi$, \\ & & about $\bfJ$ & $\phid$ for $\theta > \chi$ \\
\hline
\end{tabular}
\end{minipage}
\normalsize
\end{table*}
Given that the accretion torque may be sensitive to the relative
orientation of each possible pair of vectors in table \ref{acctable} the
full behaviour is complex, with the torque being modulated on a number of
timescales. 

To gain a little more insight consider the toy model where the torque on
the star always points along $\ndisk$, but has a magnitude that
monotonically increases with the angle between $\ndisk$ and $\bfm$.  Then
by considering the relative orientation of the vectors of table
(\ref{acctable}) it is straightforward to find a precessional phase at
which the torque, when averaged over several spin periods, pumps the
precession.  This occurs even when the wobble angle is initially zero.
However, half a precession period (i.e. $\pi / \psid$) later the relative
positions of $\bfm, \nd, \bfJ$ have changed.  Then, again when averaged
over several spin periods, it is found that the wobble is either damped (if
$\theta < \chi$) or pumped (when $\theta > \chi$), but this pumping is not
as strong as before.  This is an example of an oscillatory torque. However,
the monotonicity guarantees that the pumping over one half of a
precessional phase exceeds the damping over the next, giving rise to a
secular increase in $\theta$ which depends on the values of $\theta$ and
$\chi$.  In the limit of small wobble angles this increase is proportional
to $\theta$, giving exponential pumping.  Thus, even this simple model
reproduces a number of the features identified in the general arguments
above.

A more sophisticated model is provided by Wang \& Robnik
(1982). \nocite{wr82} Here, the torque on the star is modelled using a
magnetospheric interaction. The field outside of the star is the sum of the
usual dipole field plus an additional field due to currents induced in the
disk.  This additional field is supposed to be toroidal with respect to the
disk's symmetry axis and to be significant only near the magnetic poles of
the star.  Wang \& Robnik then calculated the torque due to the interaction
between this toroidal field and the currents internal to the star.  They
found that the torque on the star was qualitatively of the same form as the
torque that will be considered in section \ref{sect:emtorq} when we look at
electromagnetic torques on isolated pulsars.  As will be shown in detail,
such a torque leads to oscillatory pumping, and also to a secular evolution
in $\theta$, which for small wobble angles is exponential in nature.
Again, a simple model has reproduced a number of the features identified in
the general arguments.

Of course, the rather general frequency counting arguments above tell us
little about the magnitude of the modulation.  Neither do they say whether
the precession is pumped or damped.  Resolution of such issues requires a
detailed model of how the torque depends on the angles involved.
Nevertheless, they show that pumping mechanism \emph{could} apply in
accreting systems.  We will now consider the particular cases of
exponential, linear and impulsive pumping and investigate the
gravitational wave fields they would produce.

\subsection{Exponential accretion torques}

In this case equation (\ref{tdexp}) applies, with $\taup = \tauat$.  As
described in section \ref{sect:genexp}, we will divide our analysis into
two parts---one where internal dissipation is ignored, and the other where
gravitational radiation reaction is ignored. 

Begin by neglecting internal dissipation.  Then arguments identical to
those of section (\ref{sect:genexp}), save for the replacement of $\taup$
with $\tauat$, lead to the upper bounds
\[
h(\taua = \taugt) = 1.6 \times 10^{-25} \ronekpc 
\]
\be
\hspace{4cm} \rfoh^{2/3} \Medd^{1/2}
\ee
for $f < \ftheta$ and
\[
h(\taua = \taugt) = 7.4 \times 10^{-26} \ubmt \ronekpc 
\]
\be 
\hspace{4cm} \rfoh^{8/3} \Medd^{1/2}
\ee
for $f > \ftheta$.

Now neglect gravitational radiation reaction.  This leads to the  upper bounds
\[
h(\taua = \taud) = 4.3 \times 10^{-29} \ronekpc 
\]
\be
\hspace{3cm} \rfoh^{1/3} \nseven \Medd
\ee
for $f < \ftheta$ and
\[
h(\taua = \taud) = 2.0 \times 10^{-29} \ubmt \ronekpc 
\]
\be
\hspace{3cm} \rfoh^{7/3} \nseven \Medd
\ee
for $f > \ftheta$.

\begin{figure}
   \centerline{ 
   \psfig{file=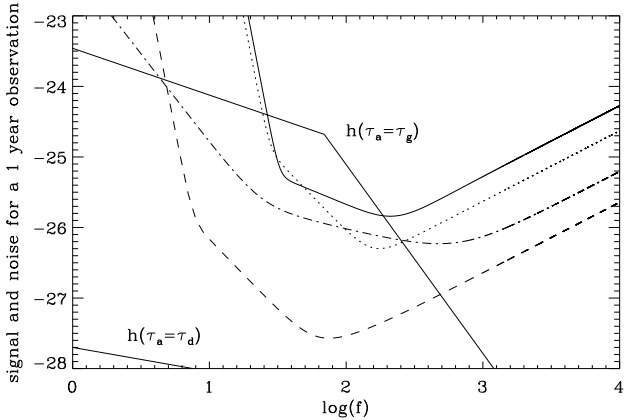,width=9cm} } 
   \caption{The maximum wave
      amplitude for an accreting star when secular accretion torques act.
      We have assumed $\ub = 10^{-3}$, $r =1$ kpc, and accretion at the
      Eddington rate.  The curve $h(\taua = \taugt)$ is the bound from
      balancing gravitational and accretion torques.  The curve $h(\taua =
      \taud)$ is the bound from balancing internal dissipation and
      accretion torques, with $n = 10^{7}$.}  
   \label{gensa}
\end{figure}

These curves are plotted in figure \ref{gensa}, with the values of $\ub,
r, \dot{M}$ and $n$ indicated by the above parameterisations.
As can be seen, at all frequencies the signal is limited by the internal
dissipation, despite the high value of $n$ assumed (three orders of
magnitude larger than the largest value estimated by Alpar \& Sauls 1988).
We therefore see that the wave amplitude due to exponential pumping by
accretion torques is almost certainly limited to undetectable values by
internal dissipation.  We will not consider this mechanism any further.

\subsection{Linear accretion torques}
\label{sect:lat}

In this case equation (\ref{tdlin}) applies, with $\taup = \tauat$.  This
will lead to upper bounds on $h$ that coincide with the low frequency
portions of the curves in figure \ref{gensa}.  Therefore internal
dissipation still limits the amplitude to undetectable values.

\subsection{Impulsive accretion torques}

We will now discuss a pumping mechanism that will apply even in the absence
of the gating described previously.  Suppose the accreting plasma is
clumped, and is not entirely confined to the plane perpendicular to the
star's rotation axis.  Various instabilities are likely to lead to such a
situation, as discussed in Lamb et al. (1985).  Indeed, the clumping of
accreting plasma is a key ingredient of the `beat-frequency' models.  These
ascribe the quasi-periodic oscillations observed in many X-ray systems to a
variability in the accretion rate.  This variability is due to a beating
between the frequency of the innermost stable orbit and the star's spin
frequency.  It therefore seems likely that the phenomenon of clumping in
accreting systems is generic.

Denote by $\Dtone$ the (average) timescale in which a clump transfers its
angular momentum to the star.  The wobble angle will change impulsively due
to the accretion of such clumps providing the transfer of angular momentum
takes place over a narrow range of rotational phase, i.e. providing $\Dtone
\ll P$, where $P$ denotes the star's spin period.  If in addition the
clumps arrive at random time intervals of average $\Dttwo$ then, providing
$\Pfp \ll \Dttwo$, a random walk in $\theta$ will occur, even though no
gating occurs.  In the absence of gravitational radiation reaction and
internal damping the wobble angle will then grow as $t^{1/2}$.  This is
clearly a less effective pumping mechanism than the linear one considered
previously, and so will not lead to interesting gravitational wave
amplitude.

\section{Electromagnetic torques}
\label{sect:emtorq}

Models of the electromagnetic torque on a spinning neutron star have split
into two classes---the Goldreich \& Julian (1969) \nocite{gj69} model where
a dipolar magnetic field embedded in a the star generates a strong electric
field at the stellar surface which rips out charged particles. These then
propagate along the magnetic field lines, forming a magnetosphere.  This
applies even when the dipole and rotation axes coincide.  The radiation the
charged particles emit then carries energy away from the star, implying the
existence of a braking torque.

The second class does not have a magnetosphere, and instead models the star
as a perfect conductor with a magnetic dipole embedded, surrounded by
vacuum  \cite{og69}. If the dipole is inclined to the rotation axis,
electromagnetic radiation at the spin frequency is emitted.  This provides
the braking torque.  Such models are clearly oversimplified.  However, the
magnetosphere models are so complicated that very little progress has been
made in understanding their details \cite{mich91}; questions such as how
the electromagnetic torque depends upon the angle between the spin axis and
dipole axis remain unanswered.  Therefore, in this section we will look at
a magnetosphere-free model in which the torque has been calculated as a
function of the spin and dipole moment vectors.  Even though the details of
such a model may prove to be incorrect, it is likely that it will reproduce
qualitatively many features of real stars, and in particular identify the
important timescales involved.

Most discussions of magnetosphere-free torques are based on the model of
Deutsch (1955), who considered a perfectly conducting sharply bounded
sphere, rigidly rotating in a vacuum.  The internal magnetic field was
assumed to be symmetric about some axis inclined to the axis of rotation.
The external magnetic and electric fields were then calculated.  These
fields were then used by Davis \& Goldstein (1970) and Michel \& Goldwire
(1970) \nocite{mg70} to calculate the torque on the star.  These authors
investigated the effect of this torque on spherical stars.  They were able
to show that the torque caused the magnetic dipole to align with the
rotation axis on the electromagnetic spin-down timescale.

Goldreich (1970) extended the analysis to examine the effect of torques on
precessing bodies.  He showed that when the free precession period was less
than the electromagnetic spin-down timescale the non-sphericity completely
altered the body's evolution---the magnetic moment no longer aligns with
the spin axis.  Instead the torque serves to either pump or damp any free
precession the body may be undergoing, depending upon whether the angle
$\chi$ between the dipole moment and the body's deformation axis is greater
or smaller than $\sin^{-1}(2/3)^{1/2} \approx 55^{\circ}$, respectively.
We will refer to this process as the \emph{Goldreich mechanism}.  This is
very different from the gravitational radiation reaction case, in which
the wobble motion is always damped \cite{cj00}.

The issue of calculating the electromagnetic torque was taken up again by
Good \& Ng (1985).  Much of their analysis was concerned with
Goldreich-Julian type magnetosphere torques.  However, in the course of
their calculation they found an error in the Deutsch fields.  When this is
taken into account the Deutsch vacuum torque $\bfT$ is given by:
\be
\label{electrofulltorque}
\bfT \equiv \bfTone + \bfTtwo =
       \frac{2 \Omega^{2}}{3c^{3}} (\bfOm \times \bfm) \times \bfm 
       - \frac{1}{5Rc^{2}}  (\bfOm  \cdot \bfm)  (\bfOm \times \bfm)
\ee 
where $\bfOm$ denotes the angular velocity, $R$ the stellar radius and
$\bfm$ the dipole moment.  (The only difference between this and the torque
as calculated from the original Deutsch paper is a factor of $-1/5$ in
$\bfTtwo$).

The first term, $\bfTone$, has components both parallel and perpendicular
to the spin axis, and scales as $\Omega^{3}$.  It is known as a
\emph{non-anomalous} torque.  It causes the energy and angular momentum of
the star to decrease.  Its component parallel to $\bfJ$ is responsible for
spin-down, while its component perpendicular to $\bfJ$ will bring about a
secular change in the wobble angle.

The second term, $\bfTtwo$, is exactly perpendicular to the angular
velocity and scales as $\Omega^{2}$.  It is known as an \emph{anomalous}
torque, and is caused by the near-zone fields.  It does not lead to a loss
in energy or angular momentum.  This torque does affect the wobble angle
and spin rate of a freely precessing star, but its effects average to zero
over one free precession period.

Following equation (\ref{electrofulltorque}) we will define two timescales
to characterise the torque.  The first corresponds to $\bfTone$ and is
simply the spin-down timescale $f/\dot{f} = 3Ic^{3}/2R^{6}\phid^{2}B^{2}$:
\be
\label{taue}
\taue = 3.64 \times 10^{3} { \rm \, yr} \rBtwelve^{2} \rfoh^{2}
        \left[\frac{1}{\sin^{2}\chi}\right].  
\ee 
The corresponding timescale on which this torque would evolve the wobble
angle is $\Icr/I$ shorter than this.  In the above formula, and those that
follow, we will place the geometric factors that follow from the particular
form of vector and scalar products of equation (\ref{electrofulltorque}) in
square brackets.  In this way it will be clear what conclusions can be
drawn if only the approximate timescales of the model hold, i.e. if in more
realistic magnetosphere models the above geometric factors were found to be
incorrect.  Such a situation would correspond to replacing the term in
square brackets with some other geometric factor.

The second timescale corresponds to $\bfTtwo$ and we define as 
\be
\tauehat = 76.2 { \rm \, yr} \rBtwelve^{2} \rfoh 
            \left[\frac{1}{\sin^{2}\chi}\right].
\ee
It is a factor $\Omega R / c$ shorter than $\taue$.  Again, the corresponding
timescale on which this torque would evolve the wobble angle is $\Icr/I$
shorter than this.

Thus the effects of the two torque terms are qualitatively different.  The
term $\bfTone$ results in a secular variation in the free precession angle
and spin rate on the electromagnetic spin-down timescale.  The term
$\bfTtwo$ produces no such secular variation.  Instead it causes
oscillations in the spin frequency and wobble angle on the (much shorter)
free precession timescale.  We will refer to this as a \emph{non-secular}
mechanism.  We will examine both torques in terms of their effect on
gravitational wave generation below.

\subsection{Non-secular electromagnetic torques}
\label{sect:nset}

The torque $\bfTtwo$ has received very little attention previously,
undoubtedly because it does not cause any secular variation in the star's
motion.  However, as can be deduced from equation (\ref{electrofulltorque})
its magnitude and orientation with respect to the reference plane varies
over one precession period, as $\bfm$ rotates with respect to this plane.
Recently Melatos (1999) made use of this torque to model spin-down
irregularities in magnetars, modelled as rigid bodies.  He considered the
case where the timescale $\tauehat$ was of similar duration to the free
precession period: $\tauehat \sim \Pfp$.  This similarity of timescales
gives a rather irregular or `bumpy' spin-down rate, which Melatos
calculated numerically.  We will take a somewhat simpler view---when
$\tauehat$ is longer than $\Pfp$ the effect of the torque will be to simply
cause a smooth variation of free precession parameters (e.g. $\theta$ and
$\phid$) calculable using perturbation theory.  Even when the two
timescales become comparable we would expect our results to apply to
order-of-magnitude accuracy.

Substitution of $\bfTtwo$ into (\ref{introthetad}) shows that for small
wobble angles, $\theta$ varies sinusoidally, giving oscillatory pumping of
the form discussed previously.  An estimate of the magnitude of this angle
follows from the timescale $\tauehat$ described above:
\be
\label{thnsestone}
\thns \sim \frac{\Pfp}{\tauehat \Icr/I}.
\ee
From equation (\ref{hosc}) this leads to a wave field
\be
\label{hestimate}
h \sim \frac{G \pi}{2 c^{4}} \frac{\Io}{r} \phid^{2} \frac{1}{\tauehat}.
\ee
This gives  
\be
\label{accuratehns}
h = 1.8 \times 10^{-28} \ronekpc  \roneyear
        \left[ \frac{3}{5 \pi} \frac{1}{\tan^{2}\chi} \right]
\ee  
where we have parameterised in terms of $\taue$ rather than $\tauehat$.  As
above the geometric factor deduced from the full perturbative calculation
has been separated from the rest of the formula.

Clearly this is extremely weak.  The amplitude increases as the
electromagnetic spin-down timescale decreases, so we should focus attention
on fast spinning strongly magnetised stars, i.e stars very soon after birth
in supernovae.  Such stars will be hot, with temperatures of order
$10^{11}$ K immediately after birth.  However, this mechanism can become
active only when the outer phase has solidified to form a crust, and so we
can apply equation (\ref{accuratehns}) only to stars which cool to the
melting temperature in a time less than $\taue$.

As described in Haensel (1997), the crust will not have a single well
defined melting temperature, as the deeper parts (with density $\sim
10^{13} \, \rm g \, cm^2$) will melt at temperatures just under $10^{10}$
K, while the outer crust (with density $\sim 10^{11} \, \rm g \, cm^2$)
will melt at temperatures just above $10^9$ K.  There is therefore some
ambiguity in what to take as an average melting temperature, with a
corresponding ambiguity in the timescale for crust formation.  For the
range of temperature identified, an upper bound would be one year, with a
lower bound much shorter than this \cite{haen97,st83}.

If the solidification timescale is indeed of order one year, equation
(\ref{accuratehns}) then leads to the tiny signal of order $10^{-29}$ for a
star born at the Galactic centre.  To be detectable over a one-year
integration the geometric factor would have to exceed $10^{2}$,
corresponding to very small dipole inclination angles, $\chi < 2^{\circ}$.

If the star's outer layers were to cool more rapidly than this we can
consider the effective amplitude found by multiplying the wave amplitude by
the square root of the number of revolutions performed in one
electromagnetic braking timescale:
\[
\label{bursthns}
h_{\rm eff} = 2.5 \times 10^{-26} \ronekpc \, \roneyear^{1/2}
\]
\be
\hspace{4cm}  
\foh^{1/2} \left[ \frac{3}{5 \pi} \frac{1}{\tan^{2}\chi} \right].
\ee
Even with fast (less than one-year) cooling, a star born at the Galactic
centre remains undetectable unless the geometric factor amplifies the
signal.  A dipole inclination of $5^{\circ}$ gives an amplification of 25.
At $100$ Hz the signal is detectable for $\taue = 1$ month, and a polar
magnetic field of $2 \times 10^{15}$ G.  Such an event could correspond to
the birth of a magnetar.  However, the event rate for such an occurrence is
very low---surely less than one a century, so that the probability of such
a source being born during the operational lifetime of Advanced LIGO is
small.

The torque $\bfTtwo$ is due to the near-zone fields and does not lead to
energy being radiated to infinity.  Therefore it is conceivable that it may
continue to act in an accreting system.  In this case the spin frequency
would be maintained by the accretion torque.  To obtain an upper bound on
the wave field at a given frequency set $\taue$ equal to its minimum value
as given by combining (\ref{faccmax}) and (\ref{taue}):
\be
\label{tauemin}
\taue = 3.4 \times 10^{7} {\rm \, yr} \foh^{1/3} \Medd^{-1}.
\ee
Substituting this into equation (\ref{accuratehns}) gives a maximum wave
amplitude of
\be
h_{\rm max}(f) = 2.4 \times 10^{-36} \ronekpc \rfoh^{1/3} \Medd.
\ee
This is tiny even in comparison to the wave amplitudes from isolated stars.
This is because in accreting systems the two factors which give large wave
amplitudes---high spin frequency and high polar magnetic field
strength---compete.  Highly magnetised stars have large Alfv\'{e}n radii
and therefore low equilibrium spin rates.

To sum up, the non-secular electromagnetic torque does not seem to be a
good pumping mechanism for gravitational wave generation.  Isolated stars
at the Galactic centre with high spin frequencies and large magnetic field
strengths can produce detectable signals if the geometric factors which
enter the calculation are favourable.  However, if born in supernovae,
their rapid spin-down rates would require extremely fast (less than
one-year) cooling of the outer phases of the star.  Accreting stars where
this near-zone torque continues to act produce even smaller gravitational
wave signals, as the requirements of high spin frequency and high magnetic
field strength oppose.

\subsection{Secular electromagnetic torques}
\label{sect:set}

Having investigated the non-secular effect of electromagnetic torques on
gravitational wave generation we will now consider the possibility that
such torques may pump precession in a secular way.  Goldreich (1970)
demonstrated that the torque $\bfTone$ of equation
(\ref{electrofulltorque}) does indeed lead to such evolution for a rigid
body.  This generalises at once to our elastic shell/fluid core model.
Substitution of $\bfTone$ into equations (\ref{introapproxphidd}) and
(\ref{introthetad}) leads to a set of two coupled differential equations
connecting $\phid$ and $\theta$.  Providing the free precession period is
less than the spin-down timescale we can average over a free precession
period to give:
\be
\label{appsecpd}
\phidd = - \alpha \phid^{3} \ssc
\ee
\be
\label{appsectd}
\thetad = \frac{I}{\Icr} \alpha \phid^{2} \theta 
          \left[ \frac{3}{2} \ssc -1 \right].
\ee
where
\be
\label{alphadef}
\alpha = \frac{2m^{2}}{3 I c^{3}}.
\ee
(Small terms proportional to $\DId /\Icr$ and $\theta^2$ have been
neglected).  Then the spin-down follows the usual power law:
\be
\label{errspindown}
\phid = \phid_{0} \left[1+\frac{2\ssc t}{\taue}\right]^{-1/2},
\ee
where $\taue$ is defined in equation (\ref{taue}).  The wobble angle
evolves according to
\be
\label{erralignment}
\theta = \theta_{0} \left(1+\frac{2\ssc t}{\taue}\right)^{\lambda}
\ee
where
\be
\lambda = \frac{I}{\Icr}
          \frac{\frac{3}{2}\ssc-1}{2\ssc}.
\ee
As can be seen, $\theta$ increases or decreases depending upon the sign of
$ \frac{3}{2} \ssc -1$, and increases most rapidly for $\chi = 90^{\circ}$,
in which case
\be
\label{thetacn}
\theta (\chi = 90^{\circ}) 
       = \theta_{0} \left(1+\frac{2t}{\taue}\right)^{I/4\Icr}.
\ee

In reality gravitational radiation reaction and internal dissipation will
act on the spinning down star also, and so we should include their effects.
Then equations (\ref{appsecpd}) and (\ref{appsectd}) acquire extra terms.
The gravitational spin-down of equation is added to
(\ref{appsecpd}), while the gravitational alignment of equation
(\ref{finaltd}) is added to (\ref{appsectd}).  The internal dissipation
alignment rate corresponding to equation (\ref{introtaud}) is added to
(\ref{appsectd}).

We will present the $\chi = 90^{\circ}$ results where the electromagnetic
pumping is greatest:
\be
\label{appcombpd}
\phidd = - \alpha \phid^{3} - \beta \phid^{5} \theta^{2}
\ee
\be
\label{appcombtd}
\thetad = \frac{1}{2} \frac{I}{\Icr}\alpha \phid^{2} \theta 
               - \beta \frac{I}{\Icr}\phid^{4} \theta
               - \frac{\DId}{\Icr} \frac{\phid}{2 \pi n} \theta
\ee
where
\be
\label{betadef}
\beta = \frac{2G}{5c^{5}I} \DId^{2}.
\ee
The terms proportional to $\alpha$ are due to electromagnetic torques,
those proportional to $\beta$ to gravitational torques, and the term
proportional to $n^{-1}$ describe internal damping.  In terms of timescales
equation (\ref{appcombtd}) would be written
\be
\label{tdtimescales}
\thetad = \theta \left[ \frac{1}{\tau_{e,\theta}} - \frac{1}{\taugt} 
        - \frac{1}{\taud} \right].
\ee
where for convenience we have introduced the electromagnetic wobble pumping
timescale
\be
\tau_{e,\theta} = 2 \taue \frac{\Icr}{I}.
\ee 
Note that the timescales in this problem are functions of the decreasing
frequency, so this equation will generally not have simple exponential
solutions.

We will apply this formalism to a young rapidly spinning down neutron star.
Its initial wobble may be due to glitches arising from the rapid spin-down
(section \ref{sect:natgl}), or due to some other process.  We will not
concern ourselves with this issue here.  We will simply examine the
evolution of the precession angle assuming that the Goldreich mechanism
acts.

Fortunately it is possible to simplify the above equations somewhat.
Firstly the gravitational radiation reaction timescale for spin-down is
much longer than for gravitational alignment, so in any situation where the
timescales in (\ref{tdtimescales}) are similar the gravitational spin-down
term is negligible, i.e. the last term of (\ref{appcombpd}) may be
neglected.  Also, the damping due to gravitational radiation reaction will
be much weaker than the damping due to internal dissipation.  Then the second
term of (\ref{appcombtd}) may be neglected.

For such a star the evolution in $\phid$ is driven only by the
electromagnetic torque and equation (\ref{errspindown}) applies.  The
evolution in $\theta$ is more complex.  Integration of (\ref{appcombtd})
depends on the variation of $\DId$ with the frequency.  We will consider
the case of Coulomb deformations, and assume $\DId \propto \phid^{2}$
(i.e. that the crust is relaxed).
We then find
\[
\label{appcombt}
\theta = \theta_{0} \left(1+\frac{2t}{\taue}\right)^{I/4\Icr}
\]
\be
\hspace{2cm}
         \exp\left\{\frac{\taue}{6\taud}\left[
          \left(1+\frac{2t}{\taue}\right)^{-1/2}-1 \right] \right\}.
\ee

\begin{figure}
   \centerline{ \psfig{file=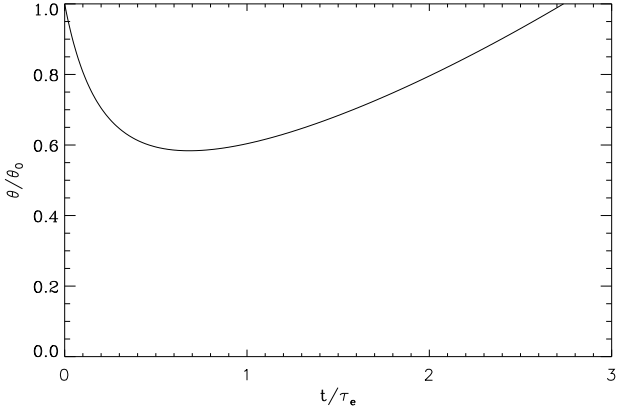,width=9cm} } \caption{ The
      evolution in the wobble angle $\theta$ under the action of combined
      internal dissipation and electromagnetic torques.  We have put $\taue =
      0.13\times \taugt$.  At early times the internal dissipation dominates
      and $\theta$ decreases exponentially.  At later times the
      electromagnetic torque dominates and $\theta$ increases as $(1+2t/
      \taue)^{I/4\Icr}$. } 
   \label{goldtheta}
\end{figure}

We can now build up a picture of the evolution of a star which is acted
upon by a secular electromagnetic torque of this form.  The scaling of the
timescales with frequency are crucial: $\taud \propto 1/\DId \phid \propto
1/\phid^{3}$, while $\taue \propto 1/\phid^{2}$.  If the star is born
spinning sufficiently fast the steeper dependence of $\taud$ on frequency
will give $\taud \ll \taue$ and the above equation reduces to
\be
\label{tearly}
\theta \approx \theta_{0} \exp{\left(-\frac{t}{\taud}\right)}.
\ee
As the star continues to spin-down the two timescales become comparable and
eventually $\theta$ will begin to increase.  At late times equation
(\ref{appcombt}) reduces to
\be
\label{tlate}
\theta \approx \left[\theta_{0} \exp{\left(-\frac{\taue}{\taud}\right)}\right]
                \left(1+\frac{2t}{\taue}\right)^{I/4\Icr}.
\ee
This is of the same form as (\ref{thetacn}), save for the extra exponential
factor.  This factor can be interpreted as the extent to which the wobble
was damped early in the star's life.  An example of this behaviour is shown
in figure \ref{goldtheta}.
Time is measured in units of $\taue$ and we have chosen $\taud =
0.13\taue$.  An example of such a star where Coulomb forces provide the
distortion could have $b=10^{-5}$, $\Icr/I = 0.1$, $B=10^{13}$ G and an
initial spin frequency of $200$ Hz.  Then $\taue = 9$ years.  The initial
exponential decrease in $\theta$, on a timescale of approximately one year,
is clearly seen.  Of course, this neglects the possibility that a CFS-type
instability (such as an r-mode) might rapidly break the star's rotation
\cite{ak01}.

In this way electromagnetic radiation reaction could provide a way of
setting young neutron stars into free precession.  Given some initial
non-zero wobble angle and the condition $\tau_{e,\theta}<\taud$, the wobble
angle increases.  Of course, this increase will end when $\theta$ reaches
the maximum value as set by the crustal breaking stress.  From that point
on the wobble angle will remain fixed at its maximum value, while the spin
frequency continues to decrease according to equation (\ref{errspindown}).
The gravitational wave amplitude is proportional to $\DId \phid^{2}$.
Given our assumption $\DId \propto \phid^{2}$, this leads to a wave
amplitude proportional to $\phid^{4}$, which in turn evolves as
$(1+2t/\taue)^{-2}$.  In other words, for all but the earliest stages of
evolution the gravitational wave amplitude decreases steadily with
time---we have been able to maintain a non-zero wobble angle only at the
expense of introducing a powerful spin-down torque.  The gravitational wave
amplitude decreases on a timescale of order $\taue$.  In section
\ref{sect:natgl} we will look at the wave field due to a population of
young \emph{unmagnetised} isolated stars, set into free precession at (or
very soon after) birth, whose precession angles then decay on the internal
damping timescale $\taud$.  Given that the wave field of the magnetised
stars considered in this section decays on the timescale
$\tau_{e,\theta}<\taud$, the magnetised stars are not as easily detectable,
and so we will not pursue them further.  Rather, in section
\ref{sect:natgl} we will consider in detail the signal strength and
population statistics of young unmagnetised stars, and find that for
realistic values of the damping parameter $n$, detection of such a
population is not likely.

To summarise, this non-secular electromagnetic pumping can increase the
wobble angle of an isolated star.  However, once the wobble angle has
increased to its maximum value the associated spin-down torques dominate
the evolution in the wave signal, with the result that the amplitude
decreases on the electromagnetic spin-down timescale.  

This motivates us to consider the possibility that secular electromagnetic
pumping remains active even in an accreting system: The accretion torque
could then maintain the star's spin frequency, while the electromagnetic
torque maintains the free precession.  It is more difficult to justify this
combining of accretion and electromagnetic torques than in the case of the
non-secular electromagnetic torque, as a secular variation in $\theta$
changes the energy and angular momentum of the star, and so must correspond
to fluxes of energy and angular momentum to infinity.  Certainly it is
clear that the exact form of the torque $\bfTone$ will no longer apply, as
its spin-down part at least will be suppressed by the accretion
environment.  We will therefore dispense with the details of the above,
i.e. the $\chi$ and numerical factors, and will assume that the star is in
spin equilibrium as described by standard accretion theory.  Then the star
is modelled by 
\begin{equation}
\label{accsts}
\thetad = \theta \left[ \frac{1}{\taue} - \frac{1}{\taugt}
                               - \frac{1}{\taud} \right],
\end{equation}
\begin{equation}
\label{spineqm}
\phidd = 0.
\end{equation}

The timescales in (\ref{accsts}) are constant by virtue of (\ref{spineqm}),
and so the solution will be exponential in form.  However, we now have to
include the details of the accretion spin-up torque, the electromagnetic
pumping torque, and internal dissipation.  In fact, it is possible to
combine formulae describing these three phenomena to obtain an upper bound
on the gravitational wave signal independent of the mechanism producing the
deformation.  Firstly, we can combine equation (\ref{taue}) for the
electromagnetic spin-down time with equation(\ref{faccmax}) which gives the
maximum spin frequency of a star with a given magnetic fields strength and
a given accretion rate.  This provides the lower bound on $\taue$ given by
equation (\ref{tauemin}).  However, for the Goldreich mechanism to be
operative we must have $\tau_{e,\theta} = 2\taue\Icr/I < \taud$.  The above
equation can then be used to give a lower bound on $\taud$, at a given spin
frequency, accretion rate and value of $\Icr/I$.  But $\taud \propto
\Icr/\DId$, and so we obtain an upper bound on $\DId$.  This immediately
become an upper bound on the gravitational wave amplitude.  Carrying out
the arithmetic we find:
\[
\hteetd = 5.0 \times 10^{-31} \Medd
              \foh^{2/3} 
\]
\be
\label{hteetdlow}
\hspace{4cm} \left(\frac{n}{10^{7}} \right) \ronekpc
\ee
for $f < \ftheta$ and
\[
\hteetd = 2.4 \times 10^{-31} \ubtmt \Medd 
\rfoh^{4/3} 
\]
\be
\label{hteetdhigh}
\hspace{3cm}
\left(\frac{n}{10^{7}} \right) \ronekpc
\ee
for $f > \ftheta$.  This is a tiny amplitude.  Even for $n= 10^{7}$ it
reaches  a maximum of only $10^{-30}$, at a frequency of $69$ Hz
for the values of $\ub$ chosen.  Therefore the gravitational wave amplitude
of accreting stars acted upon by electromagnetic torques of this form is
limited to uninteresting values by internal dissipation.

To conclude, secular electromagnetic torques are not likely to lead to
detectable levels of gravitational radiation.  In the case of isolated
stars the torque may be able to increase an initially small wobble angle to
its maximum value, but only at the expense of introducing a strong
spin-down torque.  Even if this part of the torque were to remain active in
accreting stars the wave amplitudes are uninteresting.  The bound on $h$
obtained by considering the competition between the electromagnetic torque
and the internal dissipation timescale leads to a bound on $h$ much lower
than even the Advanced LIGO sensitivity.  We will therefore not consider
the effects of electromagnetic torques any further.

\section{Natal precession and glitches}
\label{sect:natgl}

Given their violent birth in supernovae it is tempting to examine the
possibility that neutron stars are set into precession when born.  As
discussed in section \ref{sect:nset}, the high temperatures generated by
the implosion will lead to entirely fluid stars.  Only when the outer
layers have cooled will a solid crust form.  It is possible that when the
crust is in the process of solidifying a substantial excitation of a mode of
oscillation of the fluid exists.  This excitation could simply be due to
the supernova explosion itself, or could be due to a CFS-type instability.
Such a scenario has already been investigated in the context of r-modes,
where the crust formation was a hindrance to gravitational wave emission
\cite{lou00,ajks00}.  The exact outcome is not clear, particularly as the
crust will not solidify at all points simultaneously, but will form first
where the fluid velocity is smallest.  Given the complexity of the process,
the possibility of the crust/fluid core system being created in a
precessional state cannot be ruled out.

Even if the star has settled down into an axisymmetric configuration at the
time of crust solidification, it is possible that it might be set into free
precession soon after.  As noted by Lyne (1996), \nocite{lyne96} the
phenomenon of glitching---the sudden increase in spin frequency of a
pulsar---is most common in young pulsars.  This phenomenon certainly
requires a solid crust. Therefore, a young neutron star that has cooled
sufficiently will begin to glitch.  Also, some theories of glitching
associate the sudden change in spin frequency with a fracture in the
crustal structure (see Ruderman (1976, 1991a,b) and Link, Franco \& Epstein
(1998) for details).  If this change results in a shift in the principal
axis of the moment of inertia tensor the star will precess
\cite{ps73a,ps73b,lfe98}.  In this way it is possible that stars may
acquire a precessional motion very soon after birth.

We therefore wish to investigate the gravitational wave background due to a
population of young spinning-down neutron stars that were set into
precession at, or soon after, birth.  We will describe this as \emph{natal
precession}.  Although we have identified fluid modes and glitches as
possible ways of producing this natal precession, the following analysis
would apply regardless of the source of the wobble.

\subsection{Gravitational wave amplitudes}

We will not include any pumping mechanisms in our analysis, so that the
wobble angle simply decays under the combined influence of internal
dissipation and gravitational radiation reaction.  Also, we will begin by
considering stars for which electromagnetic spin-down torques are
negligible.  We would then have
\be
\label{tauglitch}
\thetad = -\theta \left[\frac{1}{\taugt} -  \frac{1}{\taud} \right]
        \equiv - \frac{\theta}{\tau}.
\ee
If we set the initial wobble angle equal to its maximum value we have
\be
\theta(t) = \thetam \exp \left( -\frac{t}{\tau} \right) 
\ee
for $t>0$ and is zero for $t<0$.  

\begin{figure}
   \centerline{ \psfig{file=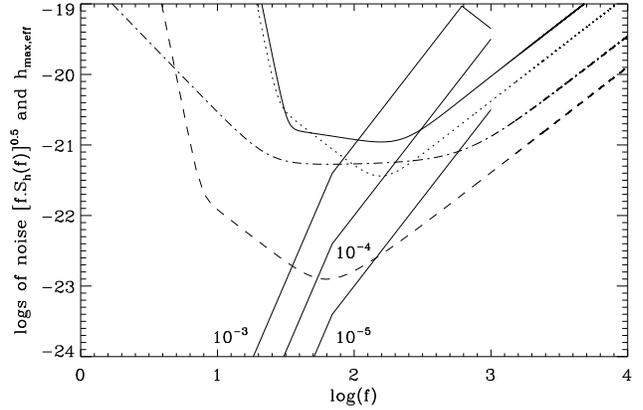,width=9cm} } \caption{The effective
      amplitude from an isolated star which initially precesses at its
      maximum wobble angle with damping due to gravitational radiation
      reaction only.
      We have put $\ub = 10^{-3}$ and $r = 1$ kpc, and have assumed that
      the matched filter accumulates signal for an interval $\tau$ defined
      in equation (\ref{tauglitch}) or for an interval of one year,
      whichever is shorter.  The deformation is due to Coulomb forces, with
      rigidity parameters $b=10^{-3}, 10^{-4}$ and $10^{-5}$ as indicated.
      The knee that appears on the $b=10^{-3}$ curve is due to the
      gravitational radiation reaction timescale falling to below one year
      at high rotation rate, limiting the signal accumulated by a matched
      filter.} 
      \label{coulglthree}
\end{figure}

\begin{figure}
   \centerline{ \psfig{file=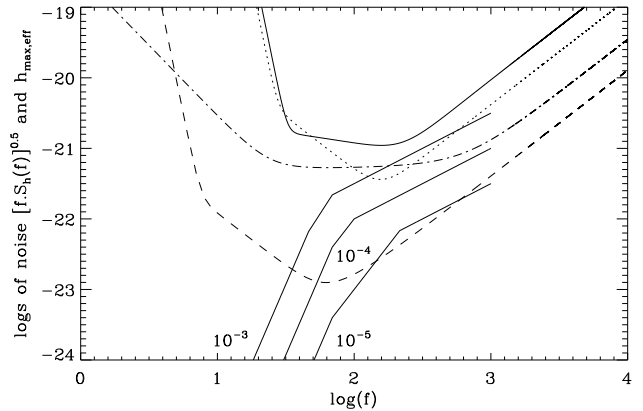,width=9cm} } \caption{The effective
      amplitude from an isolated star with the same parameters and
      assumptions as in figure (\ref{coulglthree}), except now $n=10^{4}$.
      This corresponds to the upper limit on $n$ as estimated by Alpar \&
      Sauls (1988).}
      \label{coulgltwo}
\end{figure}

In order to gain insight into the detectability of the gravitational wave
field due to such a source we shall assume that a matched filter can
accumulate signal only for an interval $\tau$ or for an interval of one
year, whichever is shorter. Figures \ref{coulglthree} and \ref{coulgltwo} show
the effective amplitude for a star deformed by crustal Coulomb forces, with
a strain angle $\ub = 10^{-3}$, at a distance 1 kpc. 
Each figure plots $h$ for three different values of the rigidity parameter:
$b=10^{-3}, 10^{-4}$ and $10^{-5}$. We have assumed a reference oblateness
equal to that of a rotating fluid, i.e. $\eo = \eom$, so that $\DId = 3
I_{\rm star} b \eom /2$.  In figure \ref{coulglthree} only the
gravitational radiation reaction term of equation (\ref{tauglitch}) is
included, i.e. this is the $n \rightarrow \infty$ limit.  Figure
\ref{coulgltwo} has $n = 10^{4}$, which is the upper bound on $n$ based on
the arguments of Alpar \& Sauls (1988).  This level of damping reduces the
signal significantly as compared to figure \ref{coulglthree} for
frequencies in excess of 100 Hz.  For example, a $b=10^{-3}$ star is now
barely detectably by a first generation interferometer, while the more
plausible $b=10^{-5}$ star is now barely detectable by an Advanced LIGO.

\subsection{Statistical arguments}

Such plots as these are useful as they show how the strength of the
internal damping affects the strength of the gravitational wave signal from
an isolated star.  However, in order to decide whether such sources are of
interest it is necessary to consider the related issues of their event rate
and distance from the Earth.  To do this we need to consider the statistics
of neutron star births.

The analogous problem of the gravitational signals due to a population of
young non-precessing \emph{triaxial} neutron stars spinning down due to
gravitational wave emission has been considered by Blandford as reported in
Thorne (1987). \nocite{thor87} Blandford made use of the following
argument: Consider a source born in a supernova explosion that is spinning
down on a timescale $\tau$.  Then there will be $\tau / \dtsn$ such sources
with age $\tau$ or younger in the Galaxy, where $\dtsn$ is the interval
between Galactic supernovae.  Blandford then modelled the Galaxy as a flat
disk of radius $R$, which allowed him to estimate the distance to the
\emph{nearest} such source.  He then combined this result with
gravitational spin-down and wave amplitude formulae from the quadrupole
formalism to show that the gravitational signal arriving at Earth from the
nearest such source is proportional to $\sqrt{\dtsn}$ and is independent of
the source's frequency or triaxial ellipticity.

We wish to make the analogous argument for a population of young isolated
precessing neutron stars.  We will follow Blandford and consider stars
where electromagnetic torques are not important.  We have already shown in
section \ref{sect:set} that electromagnetic torques can, if the
Goldreich pumping mechanism is active, cause young neutron stars to be set
into free precession, but only at the cost of introducing a powerful
spin-down torque, and so a population of unmagnetised stars is probably
more easily detectable than a population of magnetised ones.

The relevant timescale $\tau$ on which the gravitational wave signal
decreases is no longer the gravitational \emph{spin-down} timescale, but is
instead the free precession \emph{alignment} timescale, dominated in almost
all physically plausible scenarios by internal dissipation.  This means
that Blandford's triaxial result where $h$ is a function of $\dtsn$ only no
longer holds.

We will model the Galaxy as having a radius $R$ and thickness $D$, giving a
volume of order $R^{2}D$.  This contains $\tau / \dtsn$ stars of age $\tau$
or younger.  The average separation of this population of young stars is
then $(\dtsn R^{2}D/\tau)^{1/3}$.  Explicitly
\be
\label{deltar}
\Dr = 1.4 {\, \rm kpc}  \dtsnth^{1/3} \rtauot^{1/3}
\ee
where we have put $R=10$ kpc and $D=1$ kpc.  Of course, a more accurate
model would take into account the rate of star formation as a function of
Galactic position, with different rates applying in the central bulge and
spiral arms, for instance.  Nevertheless, equation (\ref{deltar})
represents a useful first approximation.

Equations (\ref{introhmtdone}) and (\ref{introhmtdtwo}) give the wave
amplitude of a precessing star in terms of $r, f$ and $\taud$.  We can
then set $r$ equal to $\Dr$ as given in equation ({\ref{deltar}). The wave
amplitude thus obtained will, subject to statistical variation, be the
field at Earth due to the closest source of age $\taud$ or less.  In full:
\[
h = 2.4 \times 10^{-30} \frac{n}{10^{4}} \foh^{4}
\]
\be
\hspace{3cm}        \rdtsnth^{1/3} \rtauot^{2/3}
\ee
for $f < \ftheta$ and
\[
h = 1.1 \times 10^{-30} \ubmt  \rfoh
\]
\be
\hspace{3cm}        \rdtsnth^{1/3} \rtauot^{2/3}
\ee
for $f > \ftheta$.  We are considering stars with $\tau$ in excess of 30
years, so these wave amplitudes can be compared against the noisecurves for
1-year matched filter integrations, e.g. the noisecurves of figure
(\ref{gensa}).  As can be seen, signals of $h = 10^{-27}$ lie above the
Advanced LIGO noisecurve for a wide range of frequencies ($25 \rightarrow
400$ Hz) which may well include the initial spin frequencies of neutron
stars.  However, we are considering a `blind search' i.e. an all-sky search
without any prior knowledge of the source's position and little idea of its
spin frequency.  In order to minimise numerically-generated false alarms we
will therefore take our detection criterion to be $h > 10^{-26}$.  Also we
will set $\ub = 10^{-3}$.  For definiteness consider the case where all
stars are born with a spin frequency $f = 100$ Hz.  If we consider the
optimistic case where internal dissipation is neglected ($n \rightarrow
\infty$)  then the above equations show that the
nearest such star is detectable for $\tau < 30$ years.  This is on the edge
of the applicability of our simple statistical model---there would only be
one such star in the Galaxy!  The probability of detecting such an object
during the lifetime of an Advanced LIGO detector would be borderline.  At
follows at once that damping parameters of less than $n = 10^{7}$ would be
unlikely to lead to a detectable population of isolated stars.

If, when the advanced detectors go on line, advances in computer power,
detector noisecurves or search algorithm design permit a more sensitive
search down to $h=10^{-27}$, then we obtain $\taud=10^{3}$ years for
detection.  There would be a population of around 30 stars of this age or
less in the Galaxy.  Equation (\ref{deltar}) shows that the closest would
be about 1 kpc from the Earth.  Figure \ref{coulglthree} shows that in the
case of Coulomb deformations such a star would have to have a rigidity
parameter of $10^{-4}$, an order of magnitude larger than the canonical
value.

We are now in a position to summarise our results.  We have considered the
wave field due to a population of isolated stars set into precession with
$\theta = \thetam$ soon after birth.  The signal following matched
filtering is shown in figures \ref{coulglthree} and \ref{coulgltwo} for the
case of Coulomb deformations with $\ub=10^{-3}$ and $r=1$ kpc.  They
suggest that even for damping as strong as $n=10^{4}$, a star with rigidity
parameter $b=10^{-4}$ is detectable out to about 1 kpc by an Advanced LIGO.
However, when the statistics of neutron star births are included it is
found that with this level of internal damping it is unlikely that, at a
given time, there exists such a precessing star in the Galaxy.  However,
the situation is very different if the internal damping is much weaker. In
the limit where it negligible, with a detection threshold of $10^{-26}$ it
is borderline as to whether or not there will exist a detectable star in
the Galaxy.  If this threshold can be decreased to $10^{-27}$, then there
could exist a detectable star as close at 1 kpc to Earth.  If deformed by
Coulomb forces, such a star would require a rigidity parameter of $10^{-4}$
at least.

\section{Collisions in dense environments}
\label{sect:cide}

The possibility of detecting gravitational waves from neutron stars set
into free precession following collisions with other stars has recently
been discussed in the literature.  de Araujo et al. (1994) \nocite{ddhc94}
first pointed out that the high stellar densities of globular clusters
could lead to a high rate of collision between neutron stars and other
stars.  Given that globular clusters contain an excess of millisecond
pulsars, the latter authors argued that if such collisions were effective
in exciting precession, interesting levels of gravitational wave emission
would occur.

This idea was taken up by Velloso et al. (1997) \nocite{vetal97} who
estimated possible gravitational wave amplitudes at Earth based on the
millisecond pulsar data for the globular clusters.  However, the formulae
employed were taken from de Araujo et al. (1994) who effectively assumed
that the stars were rigid, with an oblateness $\eom$.  This gave $h \propto
\phid^{2} \eom \theta$ and $\taugt \propto 1/ \eom^{2}$.  At a given wobble
angle and spin frequency this led to a wave amplitude too large by a factor
of $\eom / \ed \approx 1/b$ and an alignment timescale too fast by the 
factor $(\eom/\ed)^{2} \approx 1/b^{2}$.

In this section we will reconsider the issue of collision-induced
precession.  In the terminology of section \ref{sect:copm} this is an
example of an impulsive pumping mechanism.  This problem is very similar to
that considered in the last section, natal precession, as we are again
considering the wave field due to stars set into precession at positions
and times which can only be described in a statistical way, and then
spin-down and align.  Indeed, the problem again breaks down neatly into two
parts.  The first concerns identifying particular mechanisms, i.e. types of
collision, that lead to an interesting level of free precession.  The
second concerns finding an event rate for such an occurrence, so that a
statistical statement can be made regarding the likely detectability.  Note
that as the nearest cluster is of order 1 kpc from Earth, equations
(\ref{introhmtdone}) and (\ref{introhmtdtwo}) show that a safe condition
for detectability is $\taugt \la 10^{3}$ years, with the very optimistic
assumptions $n=10^{7}$ and a detection threshold of $h = 10^{-27}$.
Additionally, we see from figure \ref{coulglthree} and equation
(\ref{eq:thetamax}) that for a star with $b=10^{-3}$ and $\ub = 10^{-3}$
spinning at several 100 Hz, a wobble angle of order $10^{-3}$ is required
for a signal-to-noise of 10.  We therefore need to identify types of
collision involving at least one recycled neutron star which produce
precession angles of order $10^{-3}$ and have an event rate of order
$1/1000 \, \rm yr^{-1}$.  These arguments are somewhat simpler than those
used in the last section, but will suffice for this section, as we will
find very small event rates for collision, so that it will be clear that we
can rule out collisions as a mechanism for generating detectable
gravitational waves.

\subsection{Collisions of neutron stars with non-compact stars}

We begin by considering the collision of a neutron star with a non-compact
star.  Such collisions have been modelled extensively, e.g. Davis, Benz and
Hills (1992), \nocite{dbh92} who model neutron star-main sequence star
collisions, and also neutron star-red giant collisions.  In the main
sequence star case they find that a system consisting of a neutron star
surrounded by a thick accretion disk is formed when the separation at
periastron is $\la$ 1.75 times the main sequence star's radius.  In the
red giant case they find that a common-envelope system is formed when the
separation at periastron is $\la$ 2.5 times the red giant radius.  For
periastron separations significantly greater than these values they find
that the perturbation of the non-compact star is minimal.

Despite the violent effect such near-body encounters have on the
non-compact star it is difficult to see how the neutron star would be set
into free precession by such a collision.  As will be shown in section
\ref{sect:nsnse}, the gravitational tidal torque on the neutron star due
to the non-compact star is negligible.  This leaves only the material
torque on the neutron star, which will be determined by accretion flow onto
its surface.  This will be described by the standard theory, regardless of
the unusual source of the accreting material.  The accretion rate will be
limited to the Eddington value in the usual manner, so the torque will not
be impulsive.

It follows that although collisions between neutron stars and non-compact
stars are important in terms of the population evolution of globular
clusters, they are not of use as a mechanism for free precession
gravitational wave production as it is impossible to identify a way in
which the collision would set the star into free precession.

\subsection{Neutron star-neutron star encounters}
\label{sect:nsnse}
 
We will now consider encounters between two neutron stars.  Clearly, if a
direct collision were to occur, free precession would be the last
gravitational wave mechanism that we would wish to consider.  We will
therefore model a \emph{near} collision, where both gravitational and
magnetic effects will come into play, but there is no direct mechanical
contact between the stars.  We will begin by considering the gravitational
interaction in a simple Newtonian way.  Suppose one star has a spin-angular
momentum $\bfJ$ and centrifugal deformation $\DIOm$.  Then the other star
will exert a torque on this bulge, causing a forced precession.  (It is
this process that is responsible for the Earth's (forced) `precession of
the equinoxes' on a 26,000 year timescale, as the Sun and Moon exert a
torque on the Earth's centrifugal deformation).  The magnitude of the
torque acting on the star is given by \cite{gold80}:
\be
\label{forcedptorque}
T = \frac{3GM}{2r^{3}} \DIOm \sin 2 \beta,
\ee
where $\beta$ is the angle that the angular momentum of this star makes
with the normal to the plane in which the stars move, $M$ is the mass of
the other star and $r$ denotes the stars' separation.  This torque acts
perpendicular to the plane containing the stars and $\bfJ$.  We can write
the quantity $\DIOm$ in terms of the rotation frequency $\Omega$ using
equation (\ref{epsilonfluid}):
\be
\label{forcedpob}
\DIOm = I \frac{3}{2} \frac{\Omega^{2} R^{3}}{GM}.
\ee
This gives
\be
\label{forcedpttwo}
T = \frac{9}{4r^{3}} I \Omega^{2} R^{3} \sin 2 \beta.
\ee 

As the two stars approach one another this torque will grow and change in
orientation.  If the stars pass very close to one another the steep
$r^{-3}$ factor will give rise to an almost impulsive torque, acting when
the stars are at and close to periastron, where the $r^{-3}$ factor is at a
maximum.  If the interval in which the bulk of the angular momentum
transfer takes place is much less than the spin period of the star the
transfer will take place at nearly constant rotational phase, i.e. nearly
constant reference plane orientation.  It follows that the star would then
be set into free precession with a wobble angle of order
\be
\delta \theta \approx  \frac{\delta J}{J} \frac{I}{\Icr} 
                   \approx \frac{T}{\Icr \Omega^{2}}.
\ee
Inserting $T$ as given by equation (\ref{forcedpttwo}) then gives
\be
\label{thetaforcedp}
\delta \theta \approx \frac{9}{4} 
      \left( \frac{R}{d} \right)^{3} \sin 2 \beta \frac{I}{\Icr},
\ee
where $d$ denotes the separation at periastron.

We therefore see that collisions capable of producing significant wobble
angles $(\theta \sim 10^{-3})$ could conceivably occur, but would require
very close encounters, with a periastron passage of no more than ten
neutron star radii: $d \sim 10R$.  Of course, during such close passages
relativistic effects will be important, e.g. Lense-Thirring precession.  We
need not pursue these here.  All that we require is an order-of-magnitude
estimate of the collision cross section for a significant interaction to
occur.  Note that equation (\ref{thetaforcedp}) immediately rules out tidal
torques during near collisions between neutron stars and main sequence
stars as a pumping mechanism, due to the large $(d \ga 10^{6} \, \rm km)$
periastron separation.

There will be an interaction between the neutron stars' magnetic moments
also, as each dipole will tend to align with the field of the other.  The
torque on a dipole of moment $m$ in a field $B$ due to the other star is of
order $mB$.  The field $B$ scales as $r^{-3}$.  An argument analogous to
the above gravitational one then leads to a wobble angle
\be
\delta \theta \approx 10^{-22} \left( \frac{B_{1}}{10^{9} \, \rm G} \right) 
                               \left( \frac{B_{2}}{10^{12} \, \rm G} \right) 
                               \left( \frac{100 \, \rm Hz}{f} \right)^{2} 
                               \left( \frac{10 \, \rm km}{d} \right)^{3},
\ee 
where $B_{1}$ and $B_{2}$ denote the \emph{polar} field strengths of the
two stars.  We have parameterised in terms of field strengths appropriate
for a collision between a recycled and non-recycled neutron star.  This is
much smaller than the gravitationally-induced wobble angle and need not be
considered further.

Having established that near collisions could excite free precession we must now consider an event rate for such close passages.  Of  course, given that the event rate for encounters between a neutron star and a non-compact star was low, it is clear  that the event rate for such close neutron star-neutron star encounters will be extremely low.  In fact it is straightforward to show that no such near-collisions will occur over a Hubble time, using a simple model.  Suppose there are $N$ neutron stars in a globular cluster of size $R_{gc}$. Let $v_{\infty}$ denote their average velocity when far apart.  Then in a unit time this population sweeps out an effective volume of order $N A v_{\infty}$, where $A$ is a collision cross-section.  Then the probability of a \emph{given} neutron star colliding with another in this interval is of the order of this volume divided by the globular cluster volume, i.e. of order $NAv_{\infty} / R_{gc}^{3}$.  As there are $N$ such stars the probability of \emph{any one of them} colliding with any other is then $N$ times this giving a collision rate
$N^{2} Av_{\infty} / R_{gc}^{3}$.  As there are approximately 200 globular clusters in the Galaxy we obtain a Galactic collision rate  $\nu_{\rm collision}$:
\be
\nu_{\rm collision} \approx 200  \frac{N^{2} A v_{\infty}}{R_{gc}^{3}}.
\ee
If gravitational attractions were neglected, the collision cross-section would be of order $d^{2} \sim (100 \, \rm km)^{2}$.  However, gravitational focusing will increase the effective cross section as described in Verbunt \& Hut (1987): \nocite{vh87}
\be
\label{effcs}
A \approx d^{2} \left[ 1 + \frac{2GM_{\rm total}}{v_{\infty}^{2}d} \right]
\ee
where $M_{\rm total}$ denotes the sum of the masses of the two stars.  The second term on the right hand side describes the effects of gravitational focusing.  For the case of interest it is the dominant factor.  Parameterising we find an event rate 
\[
\nu_{\rm collision} \sim 10^{-11} \, {\rm yr^{-1}}
                    \left( \frac{N}{10^{3}} \right)^{2}
                    \left( \frac{M_{\rm total}}{2.8 \, M_{\odot}} \right)
\]
\be
\hspace{2cm}
                    \left( \frac{d}{100 \rm \, km} \right)
                    \left( \frac{10 \, \rm km}{v_{\infty}} \right)
                    \left( \frac{1 \, \rm pc}{R_{gc}} \right)^{3}.
\ee
Such an event rate as this makes further comment unnecessary, save to say that we will not consider stellar collisions any further.

\section{Conclusions}
\label{sect:conc}

This paper represents a systematic analysis of the detectability of
gravitational waves from freely precessing neutron stars.  It is based upon
a model commonly employed to describe the Earth's own motion.  Explicitly,
the neutron star has been modelled as an elastic shell with a fluid core,
whose angular amplitude of free precession (the \emph{wobble angle}) is
limited by its finite crustal lattice strength.  It has been known for some
time that neutron star structure may well allow detectable gravitational
wave signals at Earth, but this is the first study to attempt to identify
particular astrophysical scenarios in which such precessional motion might
be brought about and/or maintained.

Broadly speaking, our findings were pessimistic.  It proved impossible to
find astrophysical pumping mechanisms capable of giving steady
gravitational wave amplitudes detectable by an \emph{Advanced} LIGO.  This
was because of the limiting effect of dissipation mechanisms internal to
the star, even when dissipation strengths several orders-of-magnitude
smaller than theoretically estimated values were assumed.

Two qualifications are in order.  Firstly, the above conclusions were
reached for stars with oblate deformations.  In the physically less likely
case of a star with a prolate deformation, the effect of internal
dissipation would be to \emph{increase} the wobble angle.  Such a situation
is certainly interesting from the gravitational wave point of view,
although the dynamics of such a star, possibly involving crust cracking
when the wobble angle exceeds a critical value, are far from clear.
Secondly, most of the pumping mechanisms considered in this paper involved
an externally generated torque being exerted on the star.  There exists
another possibility, where the symmetry axis of the deformation shifts due
to a smooth plastic deformation of the crust.  For instance, this
deformation might be caused by accretion-induced temperature or composition
asymmetries, in the manner described by Bildsten (1998) for non-precessing
triaxial stars.  These two possible modifications to our model are
currently under investigation, to see if there may yet prove to be a way
of powering a long-lived, freely precessing gravitational wave source.

\section*{ACKNOWLEDGEMENTS}

It is a pleasure to thank Curt Cutler and Bernard Schutz for stimulating
discussions during the course of this work.  This work was supported by
PPARC grant PPA/G/1998/00606.


\begin{thebibliography}{10}




\bibitem[\protect\citename{Alpar \& Pines  }1985]{ap85}
Alpar A., Pines D., 1985, {\em Nature} {\bf 314} 334

\bibitem[\protect\citename{Alpar \& Sauls }1988]{as88}
Alpar A., Sauls J. A., 1988, {\em Ap. J.} {\bf 327} 723

\bibitem[\protect\citename{Andersson et al.\ }2000]{ajks00} 
Andersson N., Jones D. I., Kokkotas K. D., Stergioulas N., 2000, 
{\em Ap. J.} {\bf 534} L75

\bibitem[\protect\citename{Andersson \& Kokkotas }2001]{ak01}
Andersson N., Kokkotas K. D., 2001, To appear in {\em International Journal
of Modern Physics D}

\bibitem[\protect\citename{Baym \& Pines  }1971]{bp71}
Baym B., Pines D., 1971, {\em Annals of Physics} {\bf 66 } 816

\bibitem[\protect\citename{Bildsten }1998]{bild98}
Bildsten L., 1998, {\em Ap. J.} {\bf 501} L89

\bibitem[\protect\citename{Bondi \& Gold  }1955]{bg55}
Bondi H., Gold T., 1955, {\em MNRAS} {\bf 115} 41

\bibitem[\protect\citename{Cutler \& Jones }2000]{cj00}
Cutler C., Jones D. I., 2000, {\em Phys. Rev. D} {\bf 63} 024002

\bibitem[\protect\citename{Davies, Benz \& Hills  }1992]{dbh92}
Davies M. B., Benz W., Hills J. G., 1992, {\em Ap. J.} {\bf 401} 246

\bibitem[\protect\citename{Davis \& Goldstein }1970]{dg70}
Davis L., Goldstein M., 1970, {\em Ap. J.} {\bf 159} L81

\bibitem[\protect\citename{de Araujio et al.\ }1994]{ddhc94}
de Araujo J. C. N., de Freitas Pacheco J. A., Horvath J. E., Cattini M., 1994,  {\em MNRAS} {\bf 271} L31

\bibitem[\protect\citename{Deutsch }1955]{deut55}
Deutsch A. J., 1955, {\em Annales d'Astrophysique} {\bf 18(1)} 1

\bibitem[\protect\citename{Flanagan }1998]{flan98}
Flanagan \`{E}. \`{E}., 1998, in  Dadhich N., Narlikar J., eds, Gravitation and Relativity: At the turn of the Millennium, Proceedings of the 15th International Conference on General Relativity and Gravitation (GR15), IUCAA, p.177

\bibitem[\protect\citename{Goldreich  }1970]{gold70}
Goldreich P., 1970, {\em Ap. J.} {\bf 160} L11

\bibitem[\protect\citename{Goldreich \& Julian }1969]{gj69}
Goldreich P., Julian W. H.,   1969, {\em Ap. J.} {\bf 157} 869

\bibitem[\protect\citename{Goldstein  }1980]{gold80}
Goldstein H., 1980, Classical Mechanics, Addison-Wesley

\bibitem[\protect\citename{Good \& Ng }1985]{gn85}
Good M. L., Ng K. K., 1985, {\em Ap. J.} {\bf 299} 706

\bibitem[\protect\citename{Haensel }1997]{haen97}
Haensel P., 1997, in  Marck J.-A., Lasota J.-P., eds, Relativistic Gravitation and Gravitational Radiation, Les Houches 1995, p.129

\bibitem[\protect\citename{Jones }2001]{jonw00}
Jones D. I., 2000, PhD Thesis, University of Wales, Cardiff

\bibitem[\protect\citename{Jones \& Andersson }2001]{ja01}
Jones D. I., Andersson N., 2001, To appear in MNRAS

\bibitem[\protect\citename{Lamb, Shibazaki, Alpar \& Shaham }1985]{lsas85}
Lamb F. K., Shibazaki N., Alpar M. A., Shaham J., 1985, {\em Nature} {\bf
317} 681 

\bibitem[\protect\citename{Lamb, Lamb, Pines \& Shaham }1975]{llps75}
Lamb D. Q., Lamb F. K., Pines D., Shaham J., 1975, {\em Ap. J.} {\bf 198} L21

\bibitem[\protect\citename{Landau \& Lifshitz }1976]{ll76}
Landau L. D., Lifshitz E. M., 1976, Mechanics, 3rd Edition.  Butterworth-Heinemann Ltd.

\bibitem[\protect\citename{Lindblom, Owen \& Ushomirsky }2000]{lou00}
Lindblom L., Owen B. J., Ushomirsky G., 2000, {\em Phys. Rev. D} 
{\bf 62} 084030

\bibitem[\protect\citename{Link, Franco \& Epstein }1998]{lfe98}
Link B., Franco M. L., Epstein R. I., 1998, {\em Ap. J.} {\bf 508} 838

\bibitem[\protect\citename{Lyne  }1996]{lyne96}
Lyne A. G.,  1996, in Johnston S., Walker M. A., Bailes M., eds, Pulsars: Problems and Progress, ASP Conf. Proc. 105, p. 73

\bibitem[\protect\citename{Melatos }1999]{mela99}
Melatos A., 1999, {\em Ap. J.} {\bf 519} L77

\bibitem[\protect\citename{Michel }1991]{mich91}
Michel F. C., 1991, Theory of Neutron Magnetospheres.  The University of Chicago Press

\bibitem[\protect\citename{Michel \& Goldwire }1970]{mg70}
Michel F. C., Goldwire H. C., 1970, {\em Astrophysical Letters} {\bf 5} 21

\bibitem[\protect\citename{Misner et al.\ }1973]{mtw73}
Misner C. W., Thorne K. S., Wheeler J. A., 1973, Gravitation.  Freeman, San Francisco

\bibitem[\protect\citename{Ostriker \& Gunn }1969]{og69}
Ostriker J., Gunn J., 1969, {\em Ap. J} {\bf 157} 1395

\bibitem[\protect\citename{Owen \& Sathyaprakash}1999]{os99}
Owen B. J., Sathyaprakash B. S., 1999, {\em Phys. Rev. D} {\bf 60} 022002

\bibitem[\protect\citename{Papaloizou \& Pringle }1978]{pp78}
Papaloizou J., Pringle J. E., 1978, {\em MNRAS} {\bf 184} 501

\bibitem[\protect\citename{ Pines \& Shaham }1972a]{ps72a}
Pines D., Shaham J., 1972a, {\em Nature Physical Science} {\bf 235} 43 

\bibitem[\protect\citename{ Pines \& Shaham  }1972b]{ps72b}
Pines D., Shaham J., 1972b, {\em Phys. Earth Planet. Interiors} {\bf 6} 103 

\bibitem[\protect\citename{ Pines \& Shaham }1973a]{ps73a}
Pines D., Shaham J., 1973a, {\em Nature Physical Science} {\bf 243} 122

\bibitem[\protect\citename{ Pines \& Shaham }1973b]{ps73b}
Pines D., Shaham J., 1973b, {\em Nature} {\bf 245} 77

\bibitem[\protect\citename{Ruderman }1976]{rude76}
Ruderman M., 1976, {\em Ap J.} {\bf 203} 213

\bibitem[\protect\citename{Ruderman }1991a]{rude91a}
Ruderman M., 1991a, {\em Ap J.} {\bf 366} 261

\bibitem[\protect\citename{Ruderman }1991b]{rude91b}
Ruderman M., 1991b, {\em Ap J.} {\bf 382} 576

\bibitem
[\protect\citename{Ruderman }1992]
{rude92}
Ruderman M., 1992,
in Pines D., Tamagaki R., Tsuruta S., eds,  
Structure and Evolution of Neutron Stars, Addison-Wesley, p. 353

\bibitem[\protect\citename{Schutz }1991]{schu91}
Schutz B. F., 1991, in Blair D., The Detection of Gravitational Radiation,
p. 406 

\bibitem[\protect\citename{Shaham }1977]{shah77}
Shaham J., 1977, {\em Ap. J.} {\bf 214} 251

\bibitem[\protect\citename{Shapiro \& Teukolsky }1983]{st83}
Shapiro S. L., Teukolsky S. A., 1983, Black Holes, White Dwarfs, and Neutron Stars, Wiley-Interscience

\bibitem[\protect\citename{Thorne }1987]{thor87}
Thorne K. S., 1987, in Hawking S. W., Israel W., eds, 300 Years of Gravitation,   Cambridge University Press, p. 330

\bibitem[\protect\citename{Ushomirsky, Cutler \& Bildsten }2000]{ucb2000}
Ushomirsky G., Cutler C., Bildsten L., 2000, {\em MNRAS} {\bf 319} 902

\bibitem[\protect\citename{Velloso et al.\  }1997]{vetal97}
Velloso W., Barone F., Calloni E., Fiore L. D., Garufi F., Grado A., Milano L., 1997, in Bassan M., Ferrari V., Francaviglia M., Fucito F, Modena I., eds, General relativity and Gravitational Physics: Proceedings of the 12th Italian Conference, World Scientific Press, p457

\bibitem[\protect\citename{Verbunt \& Hut  }1987]{vh87}
Verbunt F., Hut P., 1985,  in Helfand J., Huang J.-H., eds, The Origin and Evolution of Neutron Stars, IAU Symposium 125, p.187

\bibitem[\protect\citename{Vietri \& Stella }1998]{vs98}
Vietri M., Stella L., 1998, {\em Ap. J.} {\bf 503} 350

\bibitem[\protect\citename{Wagoner }1984]{wago84}
Wagoner R. V., 1984, {\em Ap. J. } {\bf 278 } 345 

\bibitem[\protect\citename{Wang \& Robnik }1982]{wr82}
Wang Y.-M., Robnik M., 1982, {\em Astron. Astrophys.} {\bf 107} 222

\bibitem[\protect\citename{Zimmermann }1978]{zimm78}
Zimmermann M., 1978, {\em Nature } {\bf 271 } 524

\bibitem[\protect\citename{Zimmermann \& Szedenits, Jr. }1979]{zs79}
Zimmermann M., Szedenits Jr. E., 1979, {\em Phys. Rev. D } {\bf 20 } 351 



\end{thebibliography}
\end{document}